\documentclass[reprint,aps,pre]{revtex4-2}
\usepackage{graphicx}
\usepackage{amssymb}
\usepackage{amsmath}
\usepackage{gensymb}
\usepackage{xcolor}
\usepackage{hyperref}
\usepackage[utf8]{inputenc}
\usepackage{comment}
\usepackage{lipsum}
\usepackage[normalem]{ulem}
\usepackage{orcidlink}
\usepackage{subfigure}
\usepackage[section]{placeins}
\usepackage[margin=1.5cm]{geometry}

\begin{document}

\title{Principal Component Analysis of Competing Correlations in Quarter-Filled Hubbard Models}

\author{Md Fahad Equbal$^1$ \orcidlink{0009-0004-0054-9068}}
 \email{mfequbal33@gmail.com}
\author{S R Hassan$^2$}
\email{shassan@imsc.res.in}
\author{M. A. H. Ahsan$^1$ \orcidlink{0000-0002-9870-2769}}
 \email{mahsan@jmi.ac.in}
\affiliation{$^{1}$Department of Physics, Jamia Millia Islamia (Central University), New Delhi $110025$, India}
\affiliation{$^{2}$The Institute of Mathematical Sciences, C.I.T. Campus, Chennai $600113$, India}

\date{\today}
\begin{abstract}
We present an unsupervised learning analysis of correlation hierarchies in the quarter-filled simple and extended Hubbard models by applying \emph{principal component analysis} (PCA) to exact-diagonalization (ED) data on $3\times4$ and $4\times4$ cylindrical clusters. While the non-interacting limit ($U=0$) provides a finite-size reference, increasing on-site repulsion $U$ induces localization and reorganizes the low-energy spectrum.
For the extended model, we examine moderate ($U=4$) and strong ($U=10$) coupling regimes, where conventional structure factors reveal familiar crossovers among charge, spin and local-pairing correlations. PCA of the corresponding correlation matrices captures these crossovers directly from the data, without assuming predefined order parameters by identifying charge-dominated, spin-dominated and pairing-dominated regimes through variance condensation into leading components. This establishes PCA as a transparent, model-agnostic framework for uncovering the hierarchy and competition of correlation channels in finite Hubbard clusters, providing a bridge between exact diagonalization and modern machine-learning diagnostics in strongly correlated systems.

\end{abstract}

\maketitle

\section{\label{intro}Introduction}

Understanding how charge, spin and pairing correlations compete and reorganize remains a central challenge in strongly correlated systems. The quarter-filled Hubbard and extended Hubbard models offer a minimal framework to examine this interplay \cite{Hitoshi2006,Rachel2024,Clay2019,Clay2025}. Within these models, the on-site repulsion $U$ favors electron localization and moment formation \cite{Imada1998, Maria2024, Abhirup2026}, while the nearest-neighbor (NN) repulsion $V$ promotes charge ordering and spatial modulation of the density \cite{Merino2001,Yakushi2012}. 
Consistent with these trends, in the weak-coupling regime $U$ gives rise to short-range spin correlations whereas $V$ induces charge-density modulations \cite{Arovas2022, VanDongen1994, Tam2006}. In the strong-coupling limit, $U$ leads to Mott-like localization while $V$ stabilizes charge-ordered states \cite{Georges1996, Amaricci2010, Davoudi2006}. The precise phase boundaries and the sharpness of the associated crossovers depend on the theoretical framework employed, including mean-field theory (MFT), quantum Monte Carlo (QMC), exact diagonalization (ED), density matrix renormalization group (DMRG) and dynamical mean field theory (DMFT) \cite{Onari2004, Petra2019, Vojta2001, Kennedy2025, White1989, Aleksey2026, Kundu2024, Leelin2003,Calandra2002,Girlando2014, Fahad2026, Fahad2025, Weichen2023}.
Despite extensive studies using DMRG \cite{Clay2025, Vojta2001}, QMC \cite{Kennedy2025, White1989} and cluster extensions of DMFT \cite{Aleksey2026, Kundu2024, Leelin2003,Calandra2002,Girlando2014,Clay2019}, a key question remains: \emph{how do these correlations collectively reorganize across parameter space, and can this structure be identified directly from the data without predefined order parameters?}

ED, though limited to small clusters \cite{Fahad2026, Fahad2025, Weichen2023}, provides full access to many-body correlation matrices and excitation spectra. This makes it a powerful testbed for exploring how competing channels---charge, spin and pairing---coexist and evolve. Traditional observables such as structure factors or susceptibilities project the correlation data onto individual wavevector components, highlighting contributions at specific wavevectors such as $(\pi,\pi)$, which are associated with staggered charge or spin modulations. \cite{Fahad2026, Fahad2025, Farid2025, Mona2026}. In doing so, they conceal the broader hierarchy of correlations encoded in the full dataset---that is, the relative dominance and mutual interplay of various orders like charge, spin and pairing fluctuation channels as they coexist and compete within the system.
To unravel this internal organization, a complementary and unbiased analysis framework is needed.

Here we adopt a \emph{machine-learning perspective} and employ \emph{principal component analysis} (PCA) as an unsupervised diagnostic of the correlation hierarchy \cite{Ian2016, Whu2017} in the quarter-filled Hubbard and extended Hubbard models. Applied directly to operator correlation matrices obtained from ED, PCA quantifies how variance in the correlation landscape redistributes among charge, spin and local-singlet pairing channels as $U$ and $V$ are tuned. Rather than serving as a black-box classifier, PCA here functions as a transparent probe that identifies dominant fluctuation directions, revealing charge-, spin-, and pairing-dominated regimes without assuming any order parameter symmetry. This variance-based viewpoint enables us to track how correlation weight---defined as the variance of a given fluctuation channel with respect to the total variance, as quantified by the eigenvalues of the covariance matrix---condenses or fragments across interaction regimes in an unsupervised manner.

Machine-learning tools have recently shown promise in uncovering hidden organization in correlated systems~\cite{Leiwang2016}. Unsupervised approaches such as PCA have been used to detect ordering transitions in classical spin models and quantum Monte Carlo datasets~\cite{Costa2017,Kiwata2019,Khatami2019}. In the Fermi-Hubbard model, PCA on thermal snapshots successfully captured the antiferromagnetic crossover~\cite{Khatami2019}, and related analyses have been extended to Green's functions~\cite{Munoz2023}, entanglement spectra~\cite{Hu2019}, and DMFT data~\cite{Arsenault2020}. In contrast, the present work applies PCA directly to many-body operator correlations derived from exact ground states, offering a complementary, wavefunction-based perspective on how correlations self-organize---unlike sampling-based approaches \cite{Costa2017}, this is an unsupervised, model-agnostic feature extraction rather than a data-driven procedure in the conventional sense.

\begin{figure}[h]
\centering
\subfigure[]{%
\includegraphics[scale=0.27]{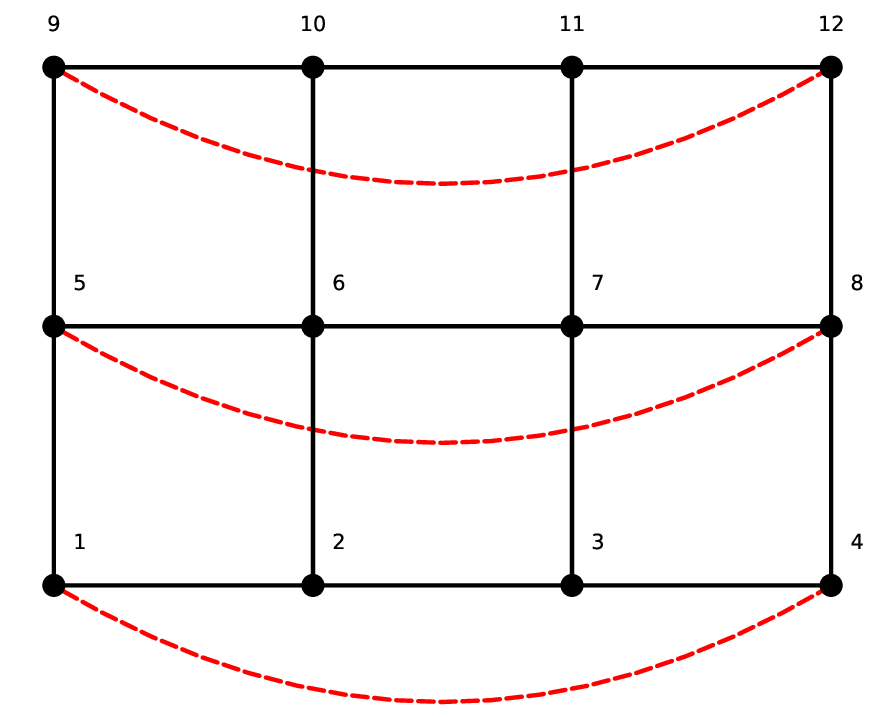}
\label{fig:3x4lat}}
\quad
\subfigure[]{%
\includegraphics[scale=0.27]{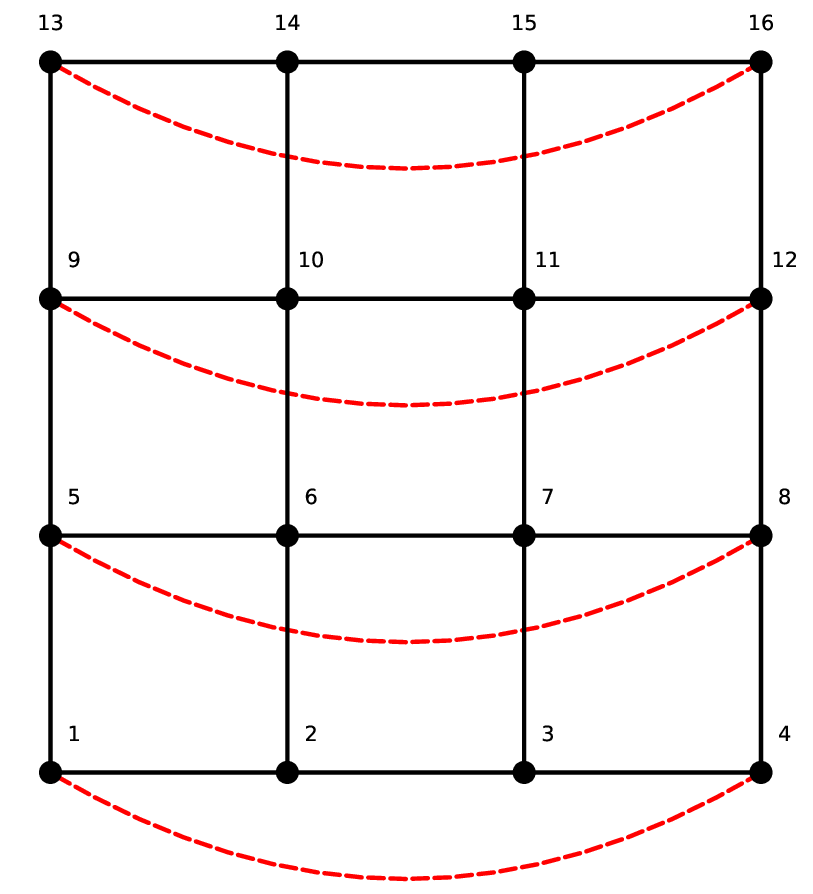}
\label{fig:4x4lat}}
\caption{Schematic of (a) $3\times4$ and (b) $4\times4$ clusters with periodic boundary conditions (PBC) along $x$ and open boundary conditions (OBC) along $y$. 
Black dots denote lattice sites, and red dashed lines represent wrap-around bonds due to PBC.}
\label{fig:latticefig}
\end{figure}

We demonstrate this framework on $3\times4$ and $4\times4$ cylindrical clusters with periodic boundary conditions along $x$ and open boundaries along $y$, shown in Fig.~\ref{fig:latticefig}. This choice facilitates comparison with DMRG studies \cite{Clay2019, Clay2025} and reduces certain finite-size effects relative to fully periodic (torus) clusters. These cluster sizes allow ED within available computational resources and permit numerically exact evaluation of correlation functions within the chosen operator basis, providing direct access to charge, spin, and pairing correlations \cite{Fahad2026, Fahad2025, Weichen2023, Mai2023, Fahadboh2025, Fahadalm2026}. Applying PCA to the resulting correlation matrices enables a unified visualization of how correlation patterns reconfigure with interaction strength, bridging conventional ED analysis with data-centric learning approaches.

The remainder of this paper is organized as follows. Section~\ref{Modmet} introduces the model, correlation functions, and PCA procedure. Section~\ref{Resdis} presents results for the simple and extended Hubbard models at moderate ($U=4$) and strong ($U=10$) couplings, emphasizing how PCA disentangles charge-, spin-, and pairing-dominated regimes. Section~\ref{summry} concludes with a summary and outlook. In what follows, we successively analyze the simple and extended Hubbard models, extracting the dominant fluctuation channels through PCA and then identifying their physical form factors from the associated eigenvectors.

\section{\label{Modmet}Model and Method}

We consider the one-band extended Hubbard model on a two-dimensional square lattice, which captures the competition between electron itinerancy, on-site Coulomb repulsion, and nearest-neighbor (NN) interactions. Its Hamiltonian reads~\cite{Micnas1990}
\begin{equation}
H = -t \sum_{\langle ij\rangle,\sigma} (c_{i\sigma}^\dagger c_{j\sigma} + h.c.)
  + U \sum_i n_{i\uparrow} n_{i\downarrow}
  + V \sum_{\langle ij\rangle} n_i n_j ,
\label{hamil}
\end{equation}
where $c_{i\sigma}^\dagger$ ($c_{i\sigma}$) creates (annihilates) an electron with spin $\sigma$ at site $i$, and $n_i = n_{i\uparrow}+n_{i\downarrow}$ is the local density operator. 
The parameters $t$, $U$, and $V$ denote the hopping amplitude, on-site, and NN Coulomb interactions, respectively; $h.c.$ indicates Hermitian conjugation. 
The model reduces to the simple Hubbard limit for $V=0$. 
Throughout this work, energies are measured in units of $t=1$, and quarter-filling corresponds to one electron per two sites ($n=0.5$). We represent number of lattice sites by $M$ and number of electrons by $N_e$. 

For the non-interacting case ($U=V=0$), the single-particle dispersion of an infinite lattice is 
$\epsilon_{\mathbf{k}} = -2t(\cos k_x + \cos k_y)$ with bandwidth $W=8t$. 
In higher dimensions, no exact analytic solution exists, and we therefore employ ED to compute ground and low-lying excited states on finite clusters. 
ED provides complete access to all correlation functions, making it ideal for testing unsupervised analysis methods despite its limited system size. 
We study $3\times4$ (12-site) and $4\times4$ (16-site) clusters with cylindrical boundary conditions (periodic along $x$, open along $y$). 
Spin-rotational symmetry is exploited to construct spin-adapted bases~\cite{SarmaAhsan1996}, which significantly reduce the Hilbert-space dimension.

The ground-state wavefunction $|G\rangle$ obtained from ED allows the evaluation of key two-point correlation functions \cite{Callaway1990}:
\begin{align}
L_{ij} &= \tfrac{1}{4}\langle G|\,(n_{i\uparrow}-n_{i\downarrow})(n_{j\uparrow}-n_{j\downarrow})\,|G\rangle, \\
D_{ij} &= \langle G|\,n_i n_j\,|G\rangle, \\
P_{ij} &= \langle G|\,c_{i\uparrow} c_{i\downarrow} c_{j\downarrow}^\dagger c_{j\uparrow}^\dagger\,|G\rangle,
\end{align}
representing spin-spin, charge-charge, and on-site singlet-pairing correlations, respectively.  
Their Fourier transforms define the corresponding structure factors,
\begin{equation}
S_X(\mathbf q) = \frac{1}{M}\sum_{ij} e^{i\mathbf q\cdot(\mathbf R_i-\mathbf R_j)} X_{ij},
\end{equation}
with $X\in\{L,D,P\}$ and characteristic momenta $\mathbf q=(\pi,\pi)$ for antiferromagnetic (AFM) or charge-density-wave (CDW) correlations, and $\mathbf q=(0,0)$ for uniform superconducting (SC) correlations. Although $\mathbf{q}=(\pi,\pi)$ corresponds to Fermi-surface nesting only at half-filling, at quarter-filling it remains a useful probe of checkerboard-type charge ordering and AFM correlations arising from effective exchange \cite{Clay2019, Merino2001, Kundu2024, Vojta2001, Calandra2002}.

While these structure factors capture conventional ordering tendencies, they provide only partial information about the full correlation landscape. To uncover the \emph{hierarchy and interplay} among charge, spin, and pairing channels, we apply PCA directly to the site-resolved correlation matrices $L_{ij}$, $D_{ij}$, and $P_{ij}$. Each matrix encodes equal-time fluctuations within its respective operator sector.

For a given correlation matrix $X \in \{L, D, P\}$ of dimension $M\times M$, the data are first centered by subtracting the mean of each row, and the covariance matrix is constructed as
\begin{equation}
C = \frac{1}{M} X^{\mathrm{T}} X .
\label{eq:covariance_matrix}
\end{equation}
Diagonalization of $C$ yields eigenvalues $\lambda_k$ and orthonormal eigenvectors $w_k$,
\begin{equation}
C w_k = \lambda_k w_k ,
\label{eq:eigenvalue_problem}
\end{equation}
where each $w_k$ defines a principal component. The normalized explained-variance ratio,
\begin{equation}
\tilde{\lambda}_k = \frac{\lambda_k}{\sum_i \lambda_i},
\label{eq:explained_variance}
\end{equation}
quantifies the fractional contribution of the k-th principal component to the total variance within that specific correlation matrix. It does not measure the absolute strength of correlations; rather, it indicates how dominant a single principal component is relative to the other components in the same channel.

In practice, the leading variance ratios $\tilde{\lambda}_k$ reveal which fluctuation channels dominate the correlation data. By tracking their evolution with interaction parameters $U$ and $V$, we can identify charge-dominated, spin-dominated, or pairing-dominated regimes in a fully unsupervised manner. This unsupervised approach thus complements traditional observables and provides a transparent framework for visualizing correlation reorganization and crossovers in finite Hubbard clusters.

\section{\label{Resdis}Results and Discussion}

We first discuss the quarter-filled \emph{simple Hubbard model} ($V=0$) on $3\times4$ and $4\times4$ cylindrical clusters, which serves as a benchmark for the interacting problem.
The goal is to understand how on-site Coulomb repulsion reorganizes charge and spin correlations
and how PCA captures these reorganizations in an unsupervised way.

\subsection{\label{sec:shm}Simple Hubbard Model ($V=0$)}

\begin{figure}[t]
\centering
\includegraphics[width=0.50\textwidth]{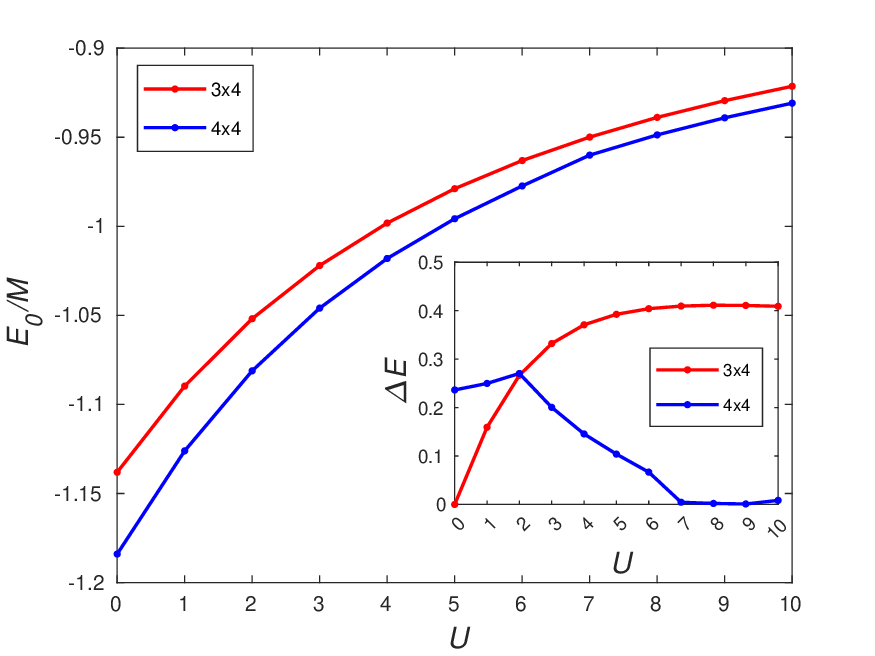}
\caption{Ground-state energy per site $E_0/M$ as a function of the on-site interaction $U$
for quarter-filled $3\times4$ (red) and $4\times4$ (blue) Hubbard clusters. The inset shows the excitation gap $\Delta E = E_1 - E_0$, revealing contrasting behaviors: a monotonically opening gap on $3\times4$ versus a collapse on $4\times4$ at large $U$.}
\label{fig:E0gapshm}
\end{figure}
We first apply PCA to the charge, spin, and pairing correlation matrices of the quarter-filled $3\times 4$ cluster.
Figure~\ref{fig:E0gapshm} shows the ground-state energy per site $E_0/M$ as a function of $U$ for both clusters, with the excitation gap $\Delta E=E_1-E_0$ shown in the inset.
As $U$ increases, $E_0/M$ becomes less negative, reflecting the reduction of double occupancy and kinetic delocalization.
The smaller $3\times4$ cluster displays a steadily \emph{increasing} $\Delta E$ that saturates near $\Delta E\simeq0.4$ for $U\gtrsim5$, indicating a stable correlated insulating state, as evidenced by the persistence of a finite excitation gap \cite{Imada1998, Limelette2003}.
By contrast, the $4\times4$ cluster exhibits a \emph{non-monotonic} $\Delta E$: the gap first increases at weak coupling and then collapses to nearly zero for $U\gtrsim7$, suggesting that the larger cluster supports many nearly degenerate low-energy states instead of developing a robust gapped phase. In finite-cluster framework, such a suppression of the excitation gap implies enhanced low-energy charge fluctuations rather than the stabilization of a well-defined insulating regime, consistent with earlier ED studies on $4\times4$ cluster \cite{Arrachea2000, Danilov2022}. This behavior indicates that despite strong on-site repulsion, the $4\times4$ cluster retains a greater degree of fluctuation-induced charge mobility than its $3\times4$ counterpart, which remains robustly gapped throughout the same regime. Such differences underscore the sensitivity of low-energy spectrum to cluster geometry and boundary conditions in finite systems. To assess the approach to the thermodynamic limit and to quantify finite-size effects, we performed a finite-size scaling analysis of the ground-state energy per site and the excitation gap. The results are presented in Appendix \ref{FSSV0}.

\begin{figure}[t]
\centering
\includegraphics[width=0.46\textwidth]{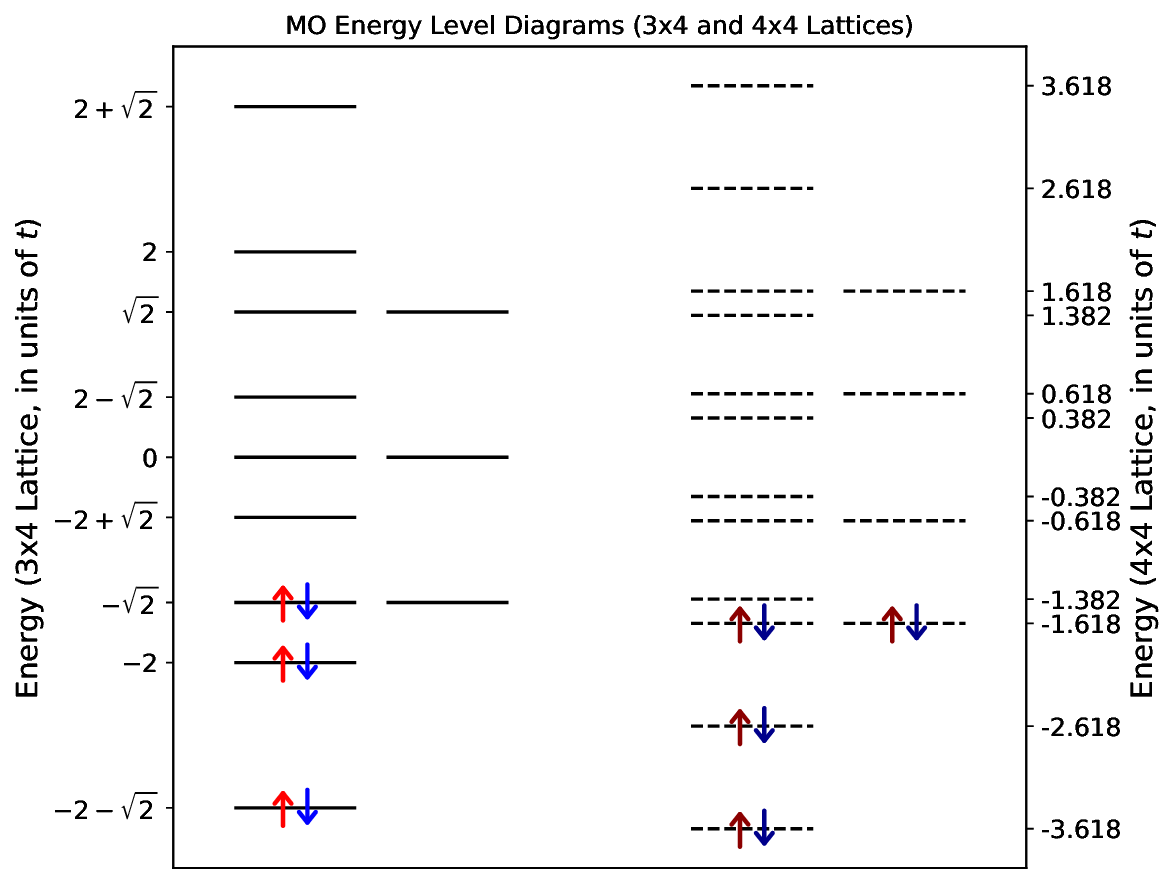}
\caption{
Single-particle molecular-orbital (MO) energy spectra of the non-interacting ($U=0$) quarter-filled Hubbard model
on $3\times4$ and $4\times4$ cylindrical lattices.
Degeneracy at the Fermi level for $3\times4$ yields a vanishing gap, while a finite highest occupied MO-lowest unoccupied MO spacing appears for $4\times4$.
}
\label{fig:mo12and16s}
\end{figure}

To rationalize these contrasting spectra, we analyze the non-interacting ($U=0$) molecular-orbital (MO) levels shown in Fig.~\ref{fig:mo12and16s}.
For $3\times4$, the highest occupied and lowest unoccupied levels are degenerate at $\epsilon_k=-\sqrt{2}$,
so the many-body gap vanishes at $U=0$.
For $4\times4$, this degeneracy is lifted, leaving a finite single-particle spacing
$\Delta\epsilon\simeq0.236t$, consistent with the finite $\Delta E$ seen in Fig.~\ref{fig:E0gapshm}.
Hence, the initial difference between the clusters originates from finite-size geometry and boundary conditions.

\begin{figure}[t]
\centering
\includegraphics[width=0.46\textwidth]{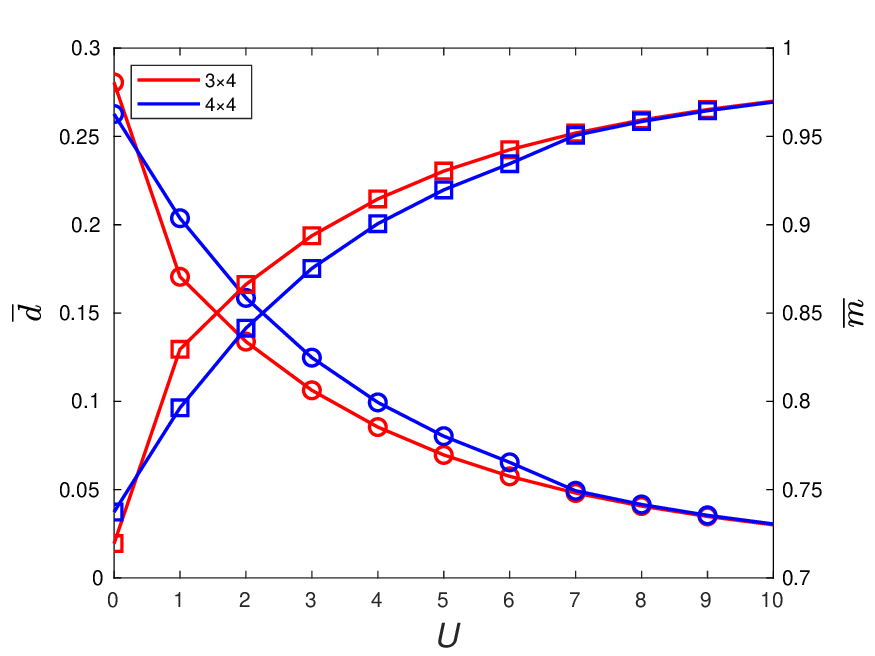}
\caption{Average double occupancy $\bar d$ (circles, left axis) and squared local moment $\bar m$ (squares, right axis) as functions of $U$ for quarter-filled $3\times4$ (red) and $4\times4$ (blue) clusters. Increasing $U$ suppresses $\bar d$ and enhances $\bar m$, reflecting progressive electron localization.}
\label{fig:dbclcmshm}
\end{figure}

The evolution of the average double occupancy $\bar d=\frac{M}{N_{e}^2}\sum_i \langle n_{i\uparrow}n_{i\downarrow}\rangle$ and local moment $\bar m=\frac{4}{N_e}\sum_i \langle(n_{i\uparrow}-n_{i\downarrow})^2\rangle$, with $U$ is shown in Fig.~\ref{fig:dbclcmshm}.
Both clusters exhibit a rapid suppression of $\bar d$ and a corresponding growth of $\bar m$, signaling the crossover from itinerant to localized behavior. Quantitatively, $\bar d$ drops by almost an order of magnitude between $U=0$ and $U=10$, while $\bar m$ approaches its saturation value, indicating fully formed local spins. In the weak to intermediate coupling regime $0<U<7$, the $3\times 4$ cluster exhibits a slightly enhanced $\bar{m}$ as compared to the $4\times 4$ cluster. For $U>7$, both clusters display identical behavior with nearly the same saturated value of $\bar{m}$. These trends indicate that, with increasing $U$, the system develops robust antiferromagnetic spin correlations as reflected in the growth of the local moment $\bar{m}$ accompanied by the suppression of charge fluctuations as evident from decreasing double occupancy $\bar{d}$.

\begin{figure}[t]
\centering
\includegraphics[width=0.50\textwidth]{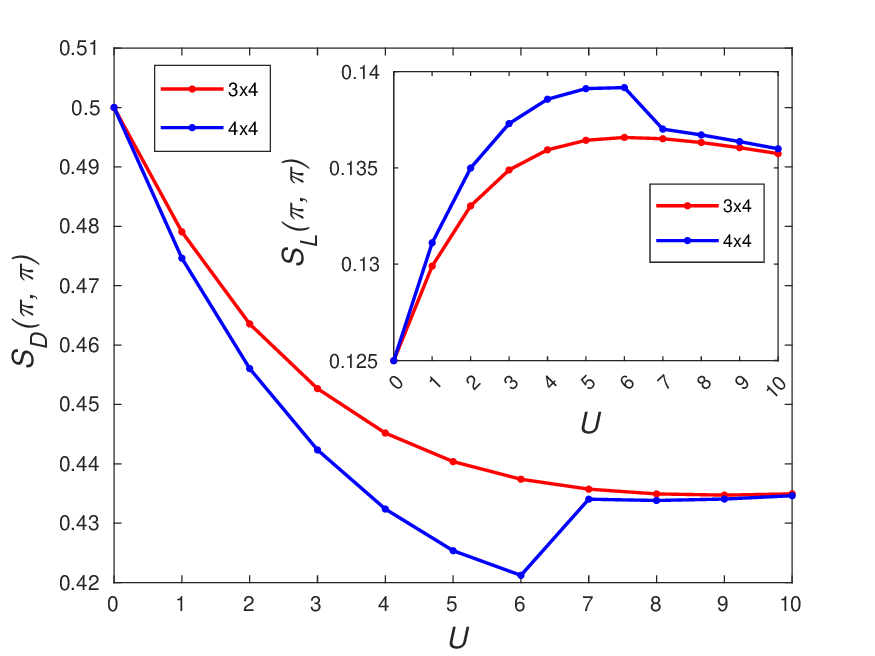}
\caption{
Charge and spin structure factors $S_D(\pi,\pi)$ and $S_L(\pi,\pi)$
versus $U$ for quarter-filled $3\times4$ and $4\times4$ Hubbard clusters.
Increasing $U$ suppresses charge modulations but strengthens antiferromagnetic spin correlations.
}
\label{fig:csfandssfshm}
\end{figure}

To quantify the evolution of charge and spin correlations, Fig.~\ref{fig:csfandssfshm} shows the corresponding charge and spin structure factors $S_D(\pi,\pi)$ and $S_L(\pi,\pi)$. The charge correlations exhibit an overall decrease with increasing $U$; for the $3\times4$ cluster this decrease is monotonic, whereas the $4\times4$ cluster displays weak non-monotonic behavior with a minimum around $U\approx6$, followed by a small increase and eventual saturation at large $U$. In contrast, the spin structure factor increases with $U$ and gradually saturates, reflecting a redistribution of correlation weight from charge to spin fluctuations. The comparatively larger values of $S_L(\pi,\pi)$ in the $4\times4$ cluster indicate more pronounced antiferromagnetic correlations, consistent with enhanced spin correlations.
Notably, in the intermediate regime $6 \lesssim U \lesssim 7$, the $4\times4$ cluster shows an enhancement in $S_D(\pi,\pi)$ accompanied by a complementary behavior in $S_L(\pi,\pi)$. These features signal a crossover from charge-density-wave (CDW) to spin-density-wave (SDW) tendencies in this parameter range. The onset of this crossover coincides with the collapse of the excitation gap (Fig.~\ref{fig:E0gapshm}) and the strengthening of magnetic correlations (Fig.~\ref{fig:dbclcmshm}), along with the emergence of nearly degenerate low-energy spin states for $U\gtrsim7$. A systematic finite-size scaling analysis of the charge structure factor $S_c(\pi,\pi)$ is discussed in Appendix \ref{FSSV0}.

While structure factors summarize ordering tendencies, they compress the full correlation landscape into a few numbers. To expose how different correlation channels reorganize collectively, we now apply principal component analysis (PCA) to the complete site-resolved correlation matrices. By tracking how the leading variance ratios evolve with $U$, we can visualize the transfer of dominance from charge to spin correlations in an unsupervised manner.

\subsection{\label{sec:pcashm}Principal Component Analysis of the Simple Hubbard Model}

\begin{figure}[t]
\centering
\includegraphics[width=0.44\textwidth]{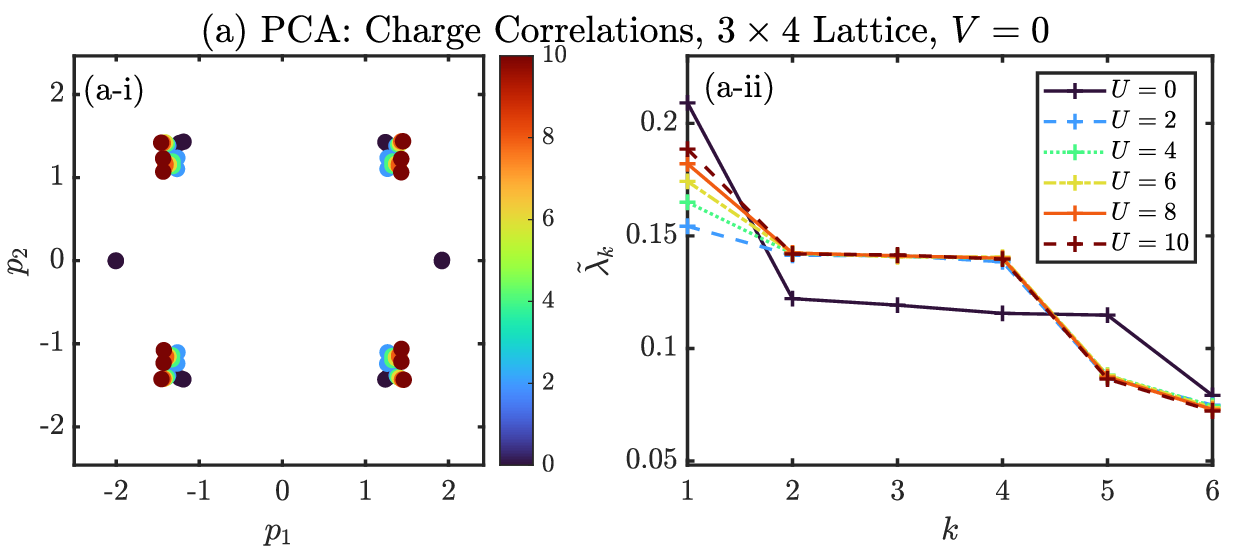}
\hspace{0.4cm}
\includegraphics[width=0.44\textwidth]{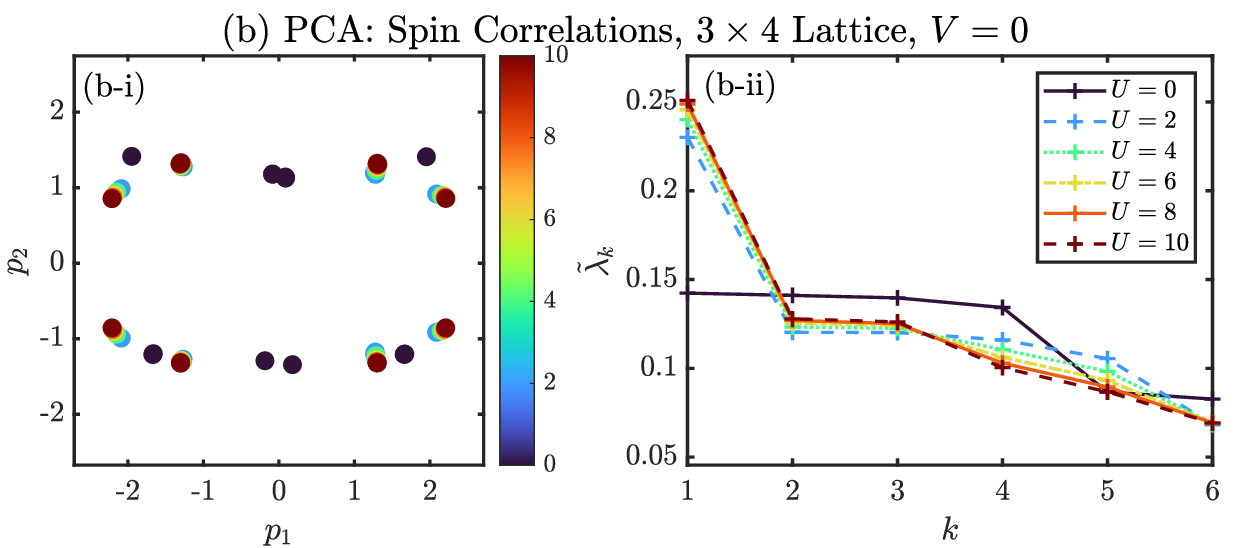}
\caption{Principal-component analysis (PCA) of (a) charge and (b) spin correlation matrices
for the quarter-filled $3\times4$ Hubbard cluster. Panels (a-i) and (b-i) show data projected onto the first two components ($p_1$, $p_2$) for $U=0,2,4,6,8,10$; colors encode $U$. Panels (a-ii) and (b-ii) display the explained-variance ratios $\tilde{\lambda}_k$ of the first six components. The diminishing $\tilde{\lambda}_1$ in charge and its growth in spin
quantify the crossover from charge- to spin-dominated fluctuations with increasing $U$.}
\label{fig:3x4pcav0}
\end{figure}

To gain an unsupervised view of how correlations reorganize with increasing $U$, we perform PCA on the full site-resolved charge and spin correlation matrices. Each correlation matrix $X$ (of dimension $M\times M$) is centered by subtracting the mean of each row, and the covariance matrix $C=(1/M)X^{\mathrm{T}}X$ is diagonalized. The eigenvalues $\lambda_k$ quantify the variance captured by the $k$-th principal component, and the normalized ratios $\tilde{\lambda}_k=\lambda_k/\sum_i\lambda_i$ measure the statistical weight of independent principal components. Tracking $\tilde{\lambda}_k$ and their corresponding eigenvectors across $U$ reveals which collective channels dominate the systems internal organization.

Figure~\ref{fig:3x4pcav0} shows the PCA of charge and spin correlations for the $3\times4$ Hubbard cluster. In the charge sector [Fig.~\ref{fig:3x4pcav0}(a-i)], data are projected onto the first two principal components ($p_1$, $p_2$), which together capture the dominant fraction of the total variance ($\approx 30-33\%$, as shown in Table \ref{tab:appD_pca_charge}, Appendix \ref{pcdata}), thereby providing a faithful low-dimensional representation of the correlation structure, with higher components not altering the qualitative clustering patterns. The resulting distribution forms four-fold symmetric clusters reflecting the underlying lattice symmetry. As $U$ increases, the variance spectrum becomes more evenly distributed and the leading eigenvalue $\tilde{\lambda}_1$ decreases from $\sim0.21$ at $U=0$ to $0.15$-$0.18$ at finite $U$ [Fig.~\ref{fig:3x4pcav0}(a-ii)]. Since $\tilde{\lambda}_1$ quantifies the contribution of the leading eigenvector---corresponding to the dominant spatial fluctuation pattern---this reduction indicates that charge correlations are no longer dominated by a single component but are instead distributed across multiple components. 

In the spin sector [Fig.~\ref{fig:3x4pcav0}(b)], $\tilde{\lambda}_1$ increases steadily from $\approx0.14$ at $U=0$ to $\approx0.25$ at $U=10$, indicating that spin correlations become increasingly dominated by a leading principal component. Correspondingly, the projection maps evolve from a relatively diffused distribution at $U=0$ to more structured clusters for finite $U$, although the separation is most pronounced between the non-interacting ($U=0$) and the interacting ($U>0$) regimes rather than between different finite-$U$ value regimes.

Taken together, the redistribution of variance across principal components---reflected in the decrease of $\tilde{\lambda}_1$ in the charge sector and its increase in the spin sector---indicates that charge correlations become more fragmented while spin correlations become more coherent. The eigenvector corresponding to the leading principal component $\tilde{\lambda}_1$ encodes information about the underlying spatial pattern, while the observed trend in the eigenvalues provide a quantitative measure of how correlation weight shifts between charge and spin channels as $U$ increases.

\begin{figure}[t]
\centering
\includegraphics[width=0.44\textwidth]{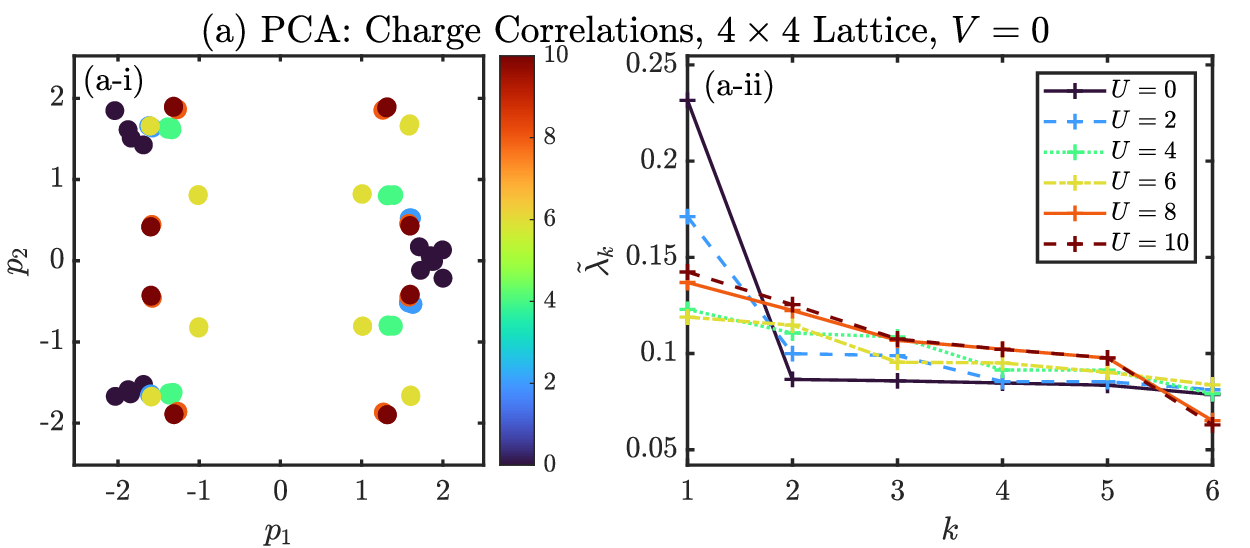}
\hspace{0.4cm}
\includegraphics[width=0.44\textwidth]{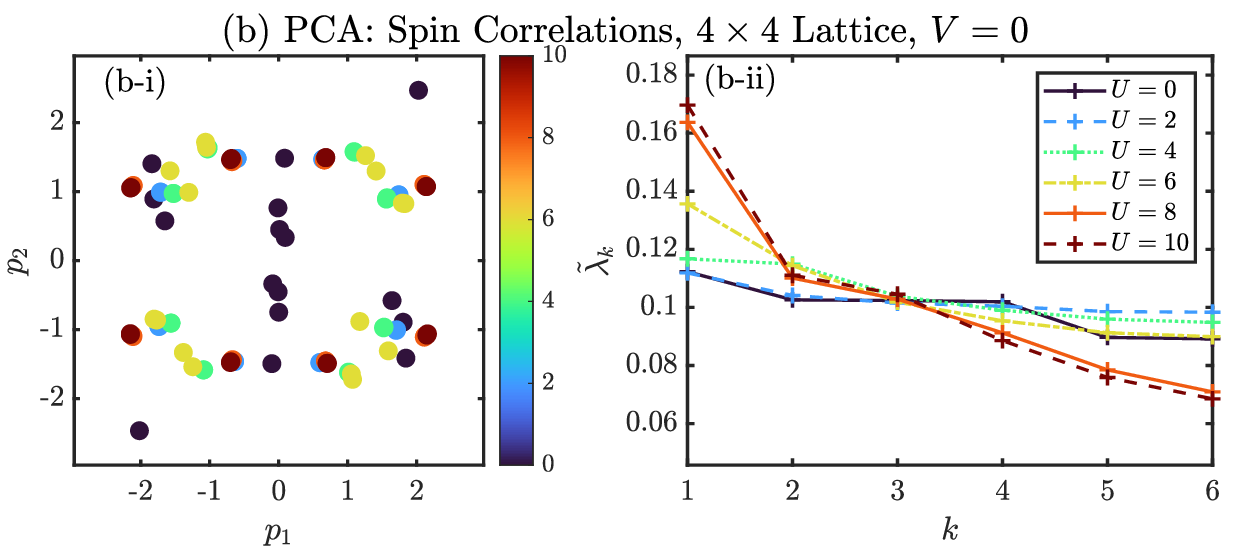}
\caption{Principal-component analysis (PCA) of (a) charge and (b) spin correlation matrices
for the quarter-filled $4\times4$ Hubbard cluster. Projections onto the first two components ($p_1$, $p_2$) for $U=0$-10 are shown in (a-i) and (b-i); the explained-variance ratios $\tilde{\lambda}_k$ are shown in (a-ii) and (b-ii). The redistribution of variance, from a dominant charge component at small $U$ to enhanced spin components at large $U$ illustrates
the interaction-driven crossover toward a spin-correlated regime.}
\label{fig:4x4pcav0}
\end{figure}

Figure~\ref{fig:4x4pcav0} presents the corresponding PCA results for the $4\times4$ lattice. In the charge sector [Fig.~\ref{fig:4x4pcav0}(a-i)], the first principal component dominates at $U=0$ ($\tilde{\lambda}_1\approx 0.23$), indicating that the leading principal component captures the largest part of the variance. As $U$ increases, $\tilde{\lambda}_1$ drops to $\sim 0.13$ while subleading principal components ($\tilde{\lambda}_2$-$\tilde{\lambda}_4$) gain comparable weights [Fig.~\ref{fig:4x4pcav0}(a-ii)]. This redistribution of weights implies that charge correlations become increasingly fragmented across several principal components. The associated eigenvectors, discussed in a later section, encode the underlying spatial patterns.

In the spin sector [Fig.~\ref{fig:4x4pcav0}(b)], $\tilde{\lambda}_1$ grows from $\sim 0.11$ to $\sim 0.17$, indicating that spin correlations become more concentrated in a leading principal component. The projection maps show increased structure for finite $U$, but the separation remains gradual.

Overall, PCA reveals an interaction-driven redistribution: charge correlations fragment while spin correlations become more coherent. This provides a quantitative, unsupervised characterization of correlation reorganization without predefined order parameters.

\subsection{\label{sec:ehmu4}Extended Hubbard Model: $U=4$}

We now introduce the nearest-neighbor Coulomb repulsion $V$ to examine how nonlocal interactions reshape the balance between charge, spin and pairing correlations in the quarter-filled extended Hubbard model. Calculations are carried out for both $3\times4$ and $4\times4$ cylindrical clusters at a representative intermediate coupling $U=4$, while $V$ is varied from $0$ to $2$ in steps of $0.2$. This parameter range is motivated by earlier studies \cite{Vojta2001, Hoang2002, Hanna2017} which show that at quarter-filling, the onset of charge-ordering tendencies occurs when $V$ becomes comparable to $U/4$. Thus, $0\le V\le2$ captures the regime where charge, spin, and pairing correlations compete most strongly.

\subsubsection{Ground-state energy and excitation gap}

\begin{figure}[t]
\centering
\includegraphics[width=0.48\textwidth]{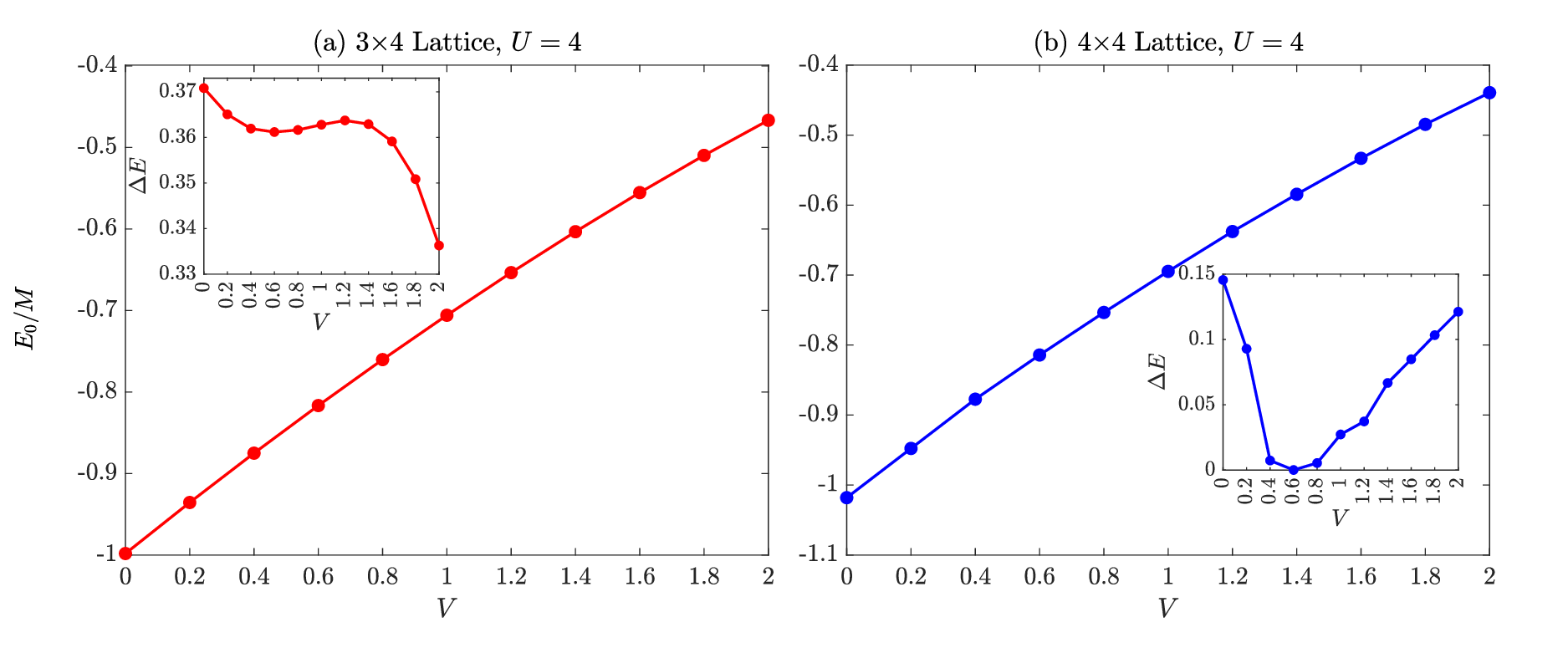}
\caption{Ground-state energy per site $E_0/M$ as a function of nearest-neighbor repulsion $V$
for quarter-filled extended Hubbard clusters at $U=4$. The inset shows the excitation gap $\Delta E=E_1-E_0$. While the $3\times4$ cluster remains insulating, the $4\times4$ system exhibits a pronounced gap collapse near $V\!\approx\!0.6$, signaling a $V$-driven reorganization of low-energy charge states.}
\label{fig:E0gapu4}
\end{figure}

Figure~\ref{fig:E0gapu4} shows the ground-state energy per site $E_0/M$ and the excitation gap $\Delta E = E_1 - E_0$ (inset) as functions of $V$. For both lattices, $E_0/M$ increases monotonically with $V$, reflecting the increasing energy cost associated with nearest-neighbor repulsion. The $3\times4$ system retains a finite and relatively weakly varying excitation gap ($\Delta E \simeq 0.35$–$0.40$), indicating a stable insulating regime across the explored range of $V$. In contrast, the $4\times4$ cluster exhibits a pronounced non-monotonic behavior: the gap decreases rapidly with increasing $V$, reaching a minimum around $V \approx 0.6$–$0.8$, and then gradually increases again for larger $V$. This suppression of the excitation gap over intermediate values of $V$ implies the presence of low-energy excitation arising from charge fluctuations. This behavior reflects crossover associated with competing interactions and enhanced charge fluctuations within a finite-size system.
Finite-size scaling behavior of the excitation gap has been examined in Appendix \ref{FSSU4}.

\subsubsection{Local observables: double occupancy and local moment}

\begin{figure}[t]
\centering
\includegraphics[width=0.48\textwidth]{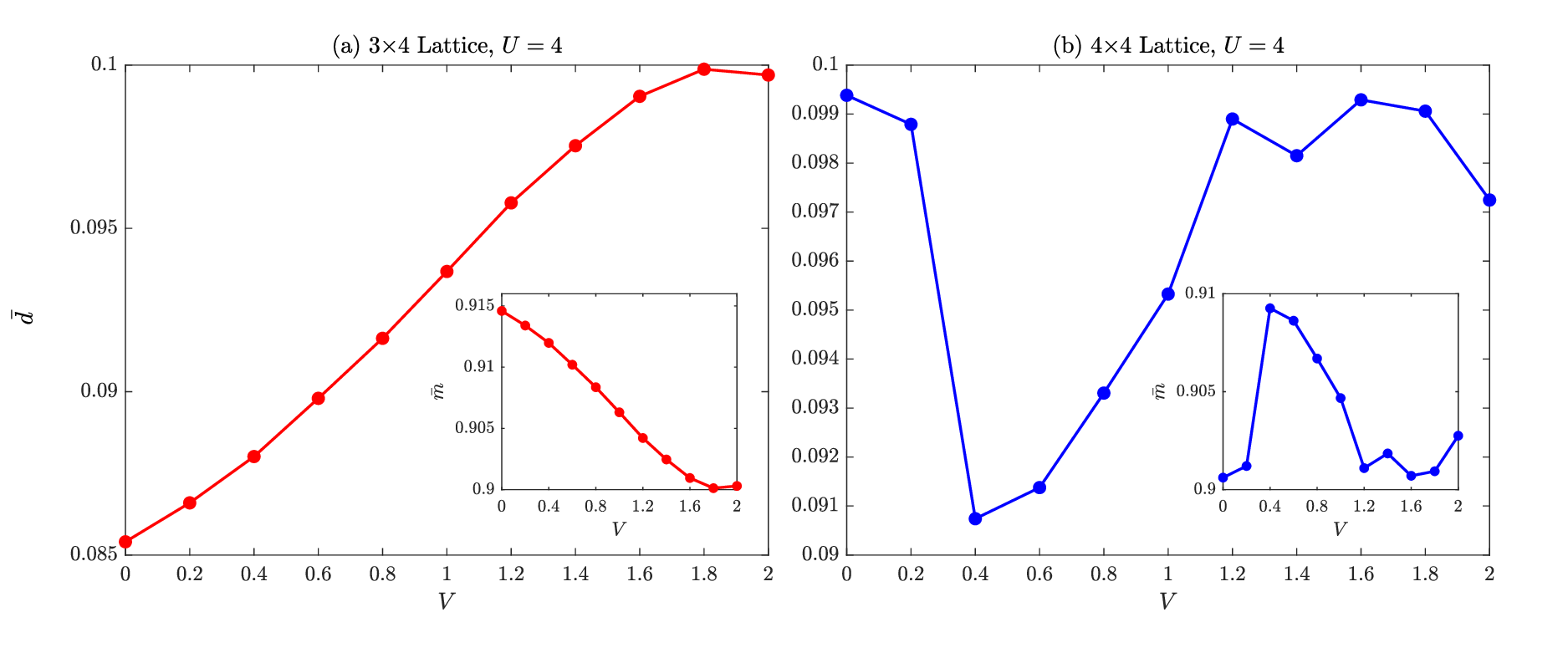}
\caption{Average double occupancy $\bar d$ and local moment $\bar m$ (inset) as functions of nearest-neighbor repulsion $V$ for the quarter-filled extended Hubbard model at $U=4$. The $3\times4$ cluster shows a smooth reduction of $\bar m$ with $V$, whereas the $4\times4$ cluster exhibits non-monotonic behavior, revealing a crossover between spin-, and charge-dominated regimes.}
\label{fig:dbclcmu4}
\end{figure}

The microscopic evolution of double occupancy $\bar d$ and local moment $\bar m$ with $V$
is summarized in Fig.~\ref{fig:dbclcmu4}. For the $3\times4$ lattice, $\bar d$ increases slightly (from $0.085$ to $0.10$) as $V$ grows, while $\bar m$ decreases marginally,
indicating that moderate nonlocal repulsion weakens spin polarization by promoting charge segregation. In contrast, the $4\times4$ lattice displays a non-monotonic evolution: $\bar d$ initially decreases up to $V\!\approx\!0.4$, suggesting enhanced localization, then rises again at larger $V$ as charge order develops. The complementary behavior of $\bar m$---peaking where $\bar d$ is minimal---signals competition between magnetic and charge degrees of freedom.
Thus, even at moderate $U=4$, the inclusion of $V$ induces rich finite-size-dependent crossovers between spin- and charge-dominated configurations.
Finite-size scaling behavior of the average double occupancy $\bar{d}$ is discussed in Appendix \ref{FSSU4}.

\subsubsection{Charge and spin correlations}

\begin{figure}[t]
\centering
\includegraphics[width=0.50\textwidth]{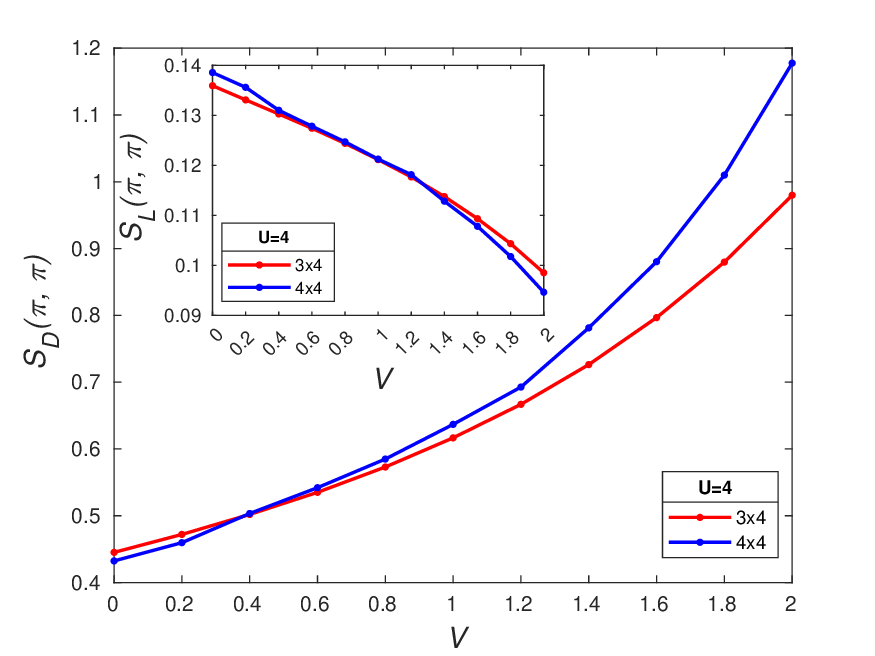}
\caption{Charge and spin structure factors $S_D(\pi,\pi)$ and $S_L(\pi,\pi)$ versus nearest-neighbor interaction $V$ at fixed $U=4$. The growth of $S_D(\pi,\pi)$ reflects enhanced CDW correlations, while the decrease of $S_L(\pi,\pi)$ indicates suppression of spin order.}
\label{fig:csfssfu4}
\end{figure}

The evolution of the charge and spin structure factors, $S_D(\pi,\pi)$ and $S_L(\pi,\pi)$, is shown in Fig.~\ref{fig:csfssfu4}. With increasing $V$, $S_D(\pi,\pi)$ rises sharply in both clusters, indicating the strengthening of CDW tendencies. The growth is steeper in the $4\times4$ lattice, consistent with its larger number of nearest-neighbor pairs and with the observed suppression of the excitation gap. Meanwhile, $S_L(\pi,\pi)$ (inset) decreases monotonically, signaling a suppression of antiferromagnetic correlations as charge fluctuations become more prominent. These trends reflect competition between the local moment formation driven by on-site interaction $U$ and the charge modulation facilitated by nearest-neighbor interaction $V$ in the extended Hubbard model \cite{Hanna2017, Hanna2018}.

\subsubsection{Local-singlet pairing correlations}

\begin{figure}[t]
\centering
\includegraphics[width=0.50\textwidth]{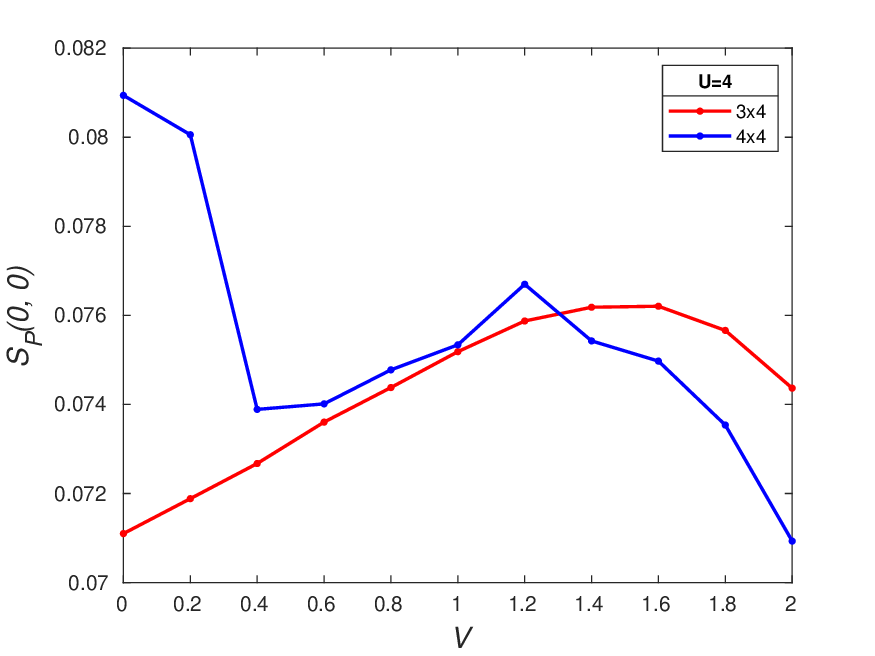}
\caption{Local-singlet pairing structure factor $S_P(0,0)$ versus $V$ for quarter-filled extended Hubbard clusters at $U=4$. Moderate $V$ enhances short-range pairing fluctuations on $3\times4$, while strong $V$ suppresses pairing on $4\times4$ as charge order becomes dominant.}
\label{fig:psfu4}
\end{figure}

To assess potential superconducting tendencies, we compute the local-singlet pairing structure factor $S_P(0,0)$, shown in Fig.~\ref{fig:psfu4}. For the $3\times4$ cluster, $S_P(0,0)$ increases slightly up to $V\simeq1.4$ before saturating, suggesting that moderate $V$ enhances local pairing fluctuations by reducing kinetic competition without establishing long-range order. In contrast, the $4\times4$ lattice displays a non-monotonic trend: a mild suppression at small $V$, a shallow maximum near $V\!\approx\!1.2$, and a clear decline for larger $V$,
consistent with the rapid growth of CDW correlations that disfavors pairing.

\subsubsection{PCA of charge, spin and pairing correlations}

\begin{figure}[t]
\centering
\includegraphics[width=0.48\textwidth]{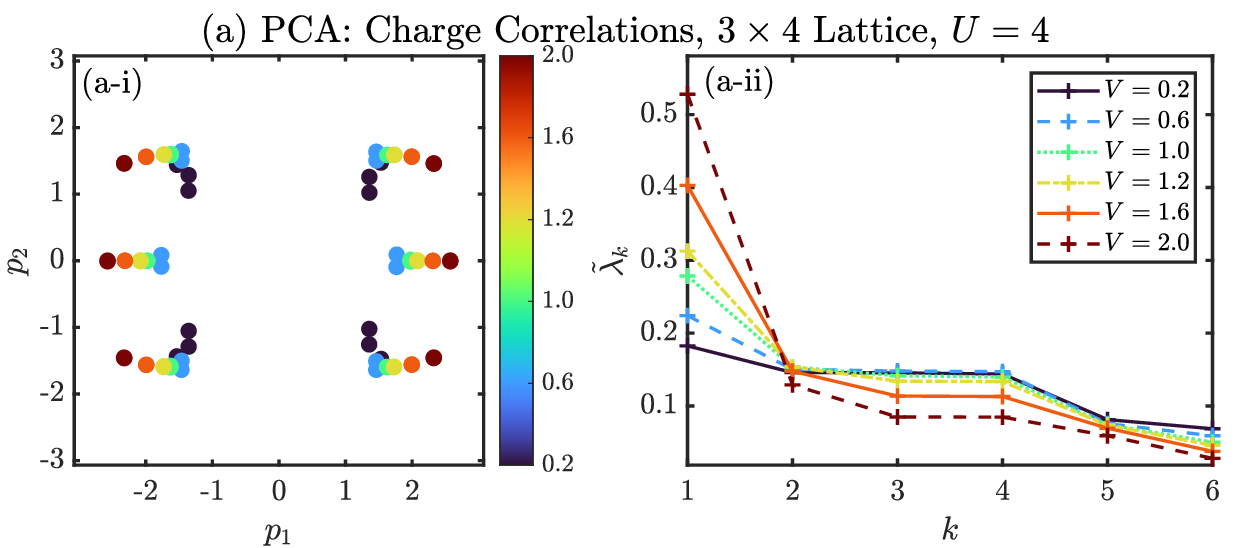}
\hspace{0.2cm}
\includegraphics[width=0.48\textwidth]{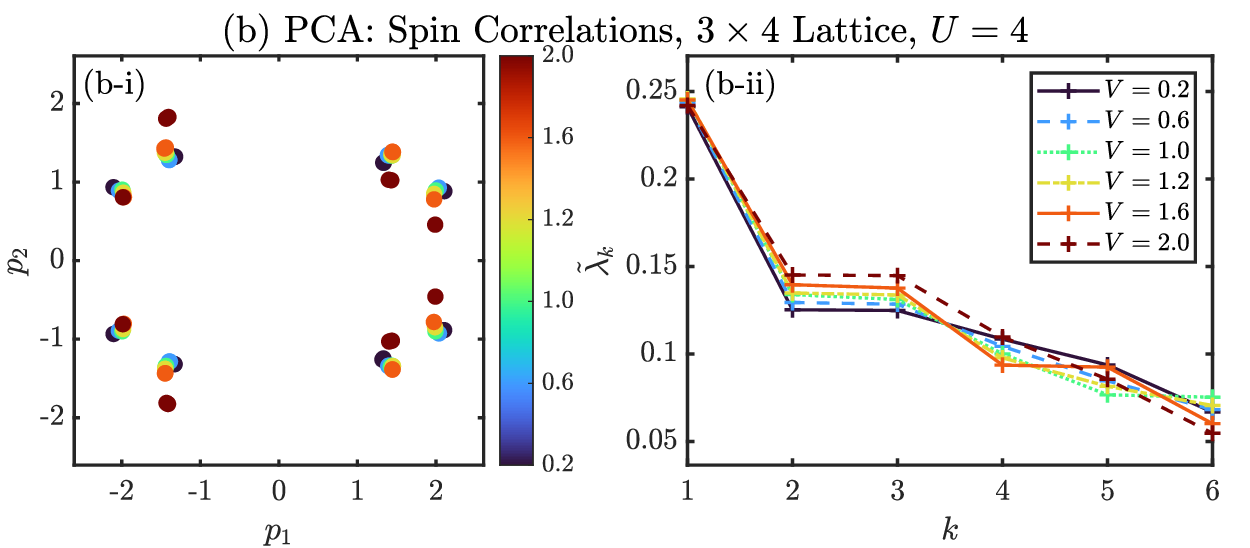}
\hspace{0.2cm}
\includegraphics[width=0.48\textwidth]{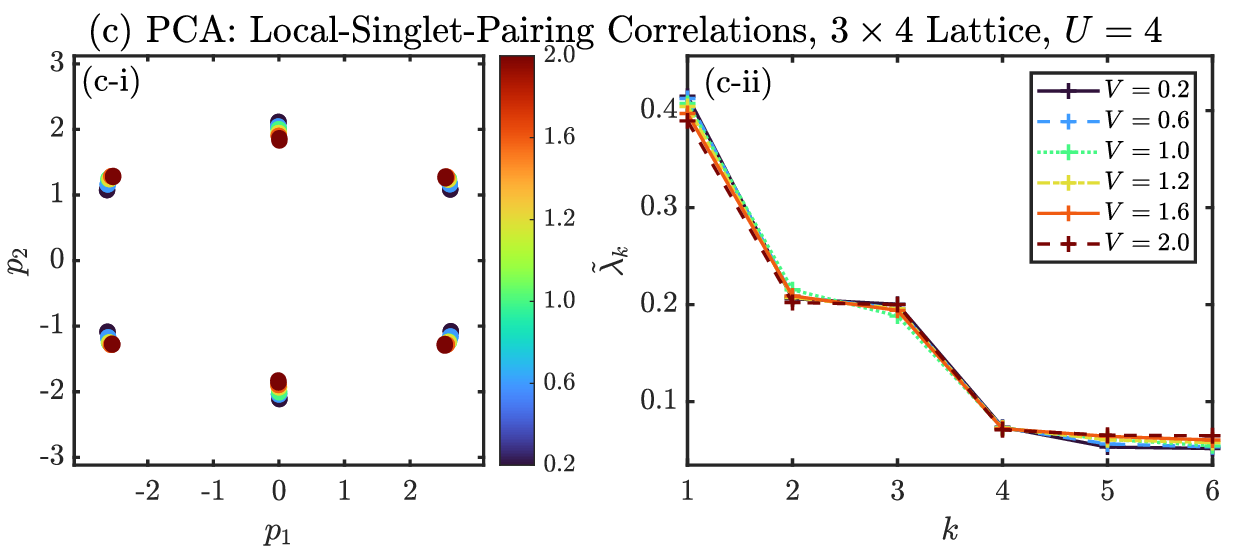}
\caption{Principal-component analysis (PCA) of (a) charge, (b) spin, and (c) local-singlet-pairing correlation matrices for the $3\times4$ cluster at $U=4$ as $V$ varies. Left panels: projections onto the first two principal components. Right panels: explained-variance spectra $\tilde{\lambda}_k$. The steady growth of $\tilde{\lambda}_1$ in the charge sector quantifies the emergence of CDW-like order with increasing $V$.}
\label{fig:3x4pcau4}
\end{figure}

To unify the above observations, we perform PCA of the charge, spin, and local-singlet-pairing correlation matrices for both cluster sizes at $U=4$. Figure~\ref{fig:3x4pcau4} displays our results for the $3\times4$ cluster.
In the charge sector [Fig. \ref{fig:3x4pcau4}(a)], projections onto the first two principal components ($p_1$, $p_2$) [Fig.~\ref{fig:3x4pcau4}(a-i)] evolve from overlapping distributions at small $V$ to a distribution of separated clusters at large $V$. This reorganization is accompanied by an increase in the leading explained-variance ratio from $\tilde{\lambda}_1\approx 0.18$ at $V=0.2$ to  $\tilde{\lambda}_1\approx 0.53$ at $V=2$ [Fig.~\ref{fig:3x4pcau4}(a-ii)] implying that charge fluctuations become progressively more concentrated in the dominant principal component.

In the spin sector [Fig. \ref{fig:3x4pcau4}(b)], projections onto the first two principal components ($p_1$, $p_2$) remain diffused and exhibit only weak evolution with $V$. Correspondingly, $\tilde{\lambda}_1$ stays nearly constant at $\approx 0.24$. This indicates that the relative importance of leading spin component within the spin correlation matrix does not change significantly.

In the pairing sector [Fig. \ref{fig:3x4pcau4}(c)], projections onto ($p_1$, $p_2$) form clusters of overlapping distributions whose arrangement shows little variation with $V$, and the leading principal component remains nearly $V$-independent with $\tilde{\lambda}_1 \approx 0.40$. This is relatively small compared to the corresponding value of the leading principal component $\tilde{\lambda}_1\approx 0.53$ in the charge sector at large $V$. Thus, the local pairing fluctuations are less dominant compared to the charge fluctuations in the corresponding PCA variance channel. However, establishing the hierarchy of various correlations, for a given value of $U$ and $V$, requires complementary observables beyond PCA.

\begin{figure}[t]
\centering
\includegraphics[width=0.48\textwidth]{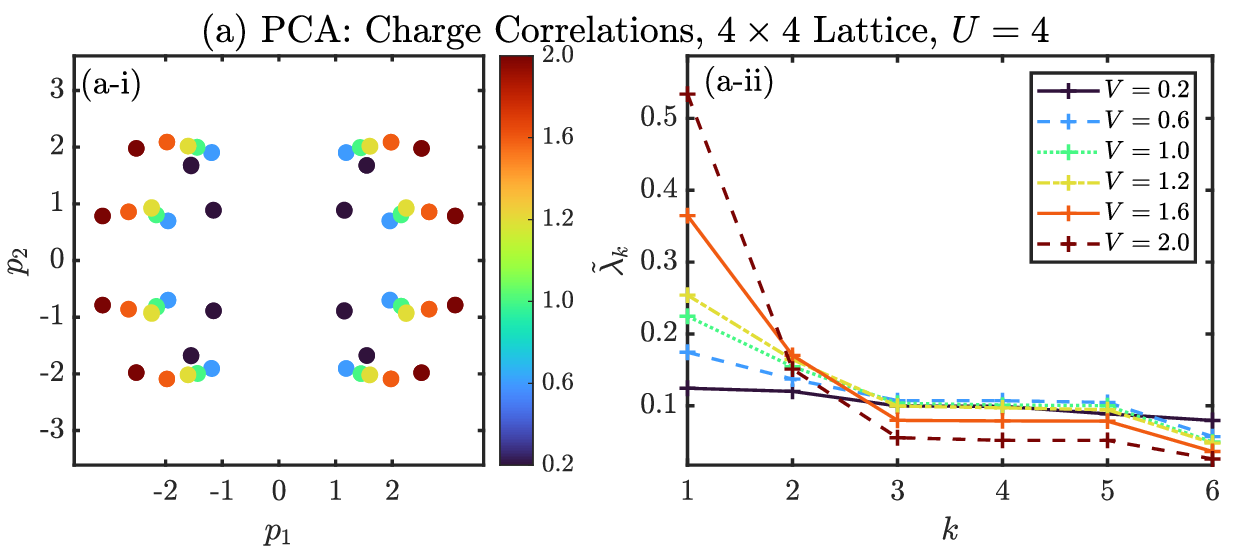}
\hspace{0.2cm}
\includegraphics[width=0.48\textwidth]{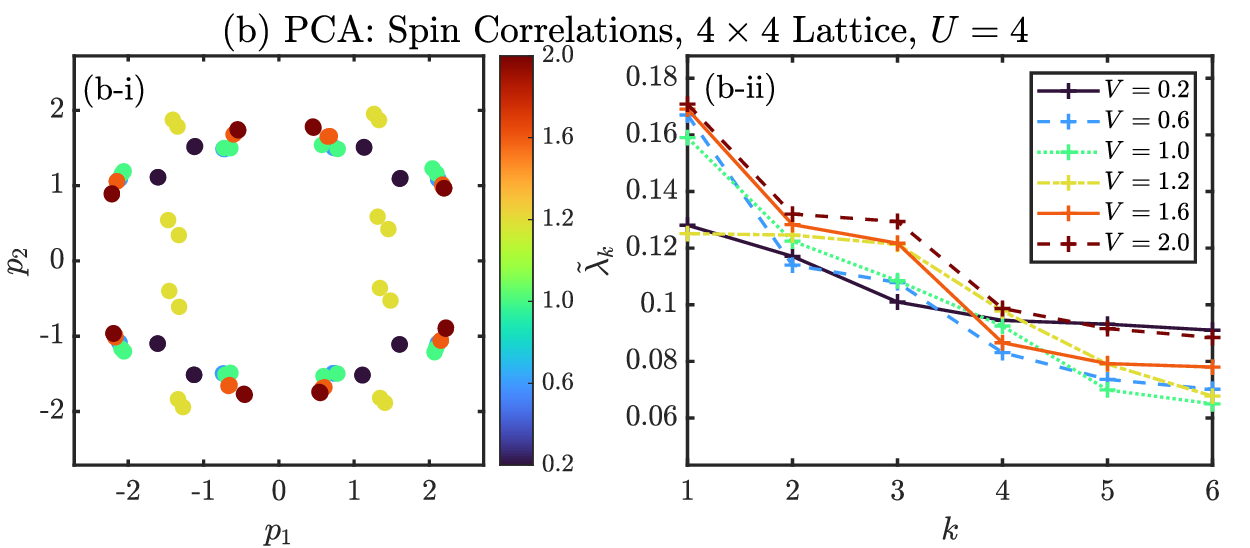}
\hspace{0.2cm}
\includegraphics[width=0.48\textwidth]{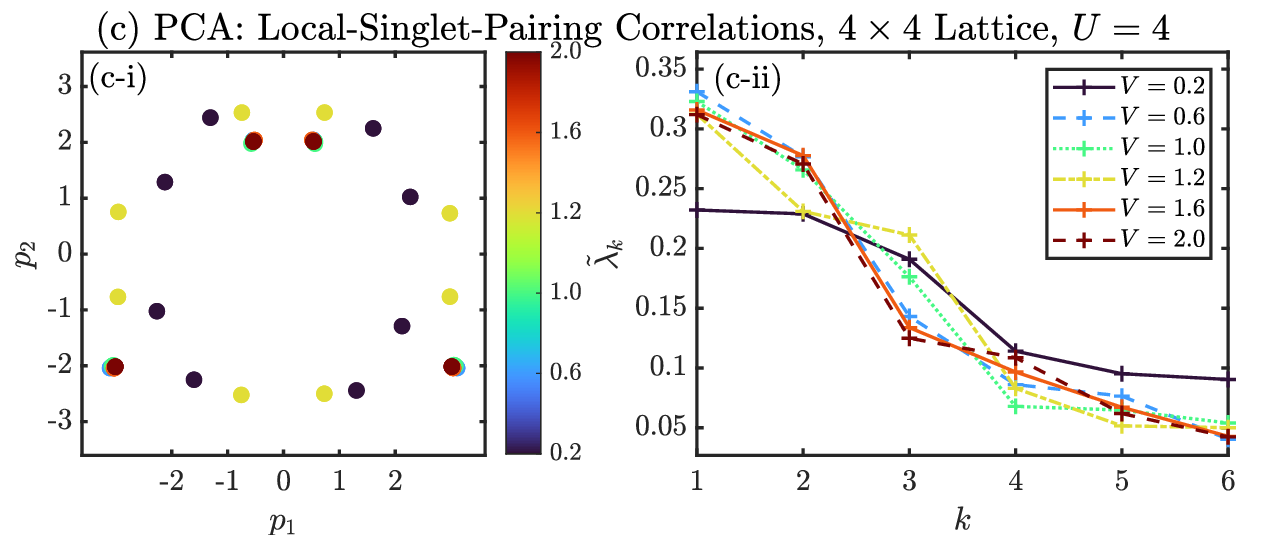}
\caption{Principal-component analysis (PCA) of (a) charge, (b) spin, and (c) local-singlet-pairing correlation matrices for the $4\times4$ cluster at $U=4$ as $V$ increases. Charge correlations become progressively dominated by a single principal mode ($\tilde{\lambda}_1\!\to\!0.5$), while spin and pairing channels remain multicomponent and less coherent,
highlighting $V$-driven charge-order enhancement.}
\label{fig:4x4pcau4}
\end{figure}

Figure \ref{fig:4x4pcau4} presents the corresponding PCA results for the $4\times4$ cluster. 
In the charge sector [Fig.~\ref{fig:4x4pcau4}(a)], projections onto ($p_1$, $p_2$) evolve from overlapping distributions at small $V$ to well-separated clusters at large $V$, indicating the emergence of structured spatial patterns. Consistently, $\tilde{\lambda}_1$ increases from $\approx 0.12$ at $V=0.2$ to $\approx 0.53$ at $V=2$, signifying that the variance becomes strongly concentrated in the dominant principal component.

In the spin sector [Fig. \ref{fig:4x4pcau4}(b)], projections onto the first two principal components ($p_1$, $p_2$) remain diffused and isotropic, with $\tilde{\lambda}_1$ varying only weakly ($\approx 0.13$-$0.17$). This indicates that no single spin component becomes dominant and the variance stays distributed across multiple components.

In the pairing sector [Fig. \ref{fig:4x4pcau4}(c)], projections onto the first two principal components occupy a broader region in the ($p_1$, $p_2$) plane and form several distinct groupings whose arrangement changes little with $V$. The explained-variance ratio spectrum shows two nearly degenerate leading components ($\tilde{\lambda}_1$, $\tilde{\lambda}_2 \approx 0.25-0.30$), whose values vary only weakly with $V$. This near-degeneracy indicates that the variance within the pairing correlation matrix is shared between two competing local pairing configurations rather than condensing into a single dominant component.

The PCA projections show that charge correlations become increasingly dominated by a single principal component as $V$ grows, while spin and pairing channels remain comparatively distributed. The real-space distribution of the leading charge-correlation eigenvector (Fig. \ref{fig:combined_charge_eigvecs}) exhibits a checkerboard modulation, characteristic of CDW order, indicating that $V$ promotes charge ordering. A large $\tilde{\lambda}_1$ alone, however, does not determine the order; rather, it is $\tilde{\lambda}_1$ together with the corresponding eigenvector that identifies the ordering tendency.

Overall, at moderate coupling $U=4$, the extended Hubbard model exhibits a clear hierarchical reorganization: as $V$ grows, the charge sector condenses into a single dominant principal component, the spin sector loses coherence, and pairing fluctuations are moderately suppressed. The PCA thus provides a transparent, quantitative visualization of how nonlocal repulsion reshapes the competing correlation landscape.

\subsection{\label{sec:ehmu10}Extended Hubbard Model: $U=10$}

In the strong-coupling limit with $U=10$, electrons are nearly localized. The nearest-neighbor interaction, considered here in the regime $0\leqslant V\leqslant U/2$, acts mainly to reorganize spatial distribution of charges, capturing crossover from spin-dominated regime at small $V$ to charge-dominated regime at large $V$ \cite{Sherman2023, Aaron2014}. 

\subsubsection{Ground-state energy and excitation gap}

\begin{figure}[t]
\centering
\includegraphics[width=0.50\textwidth]{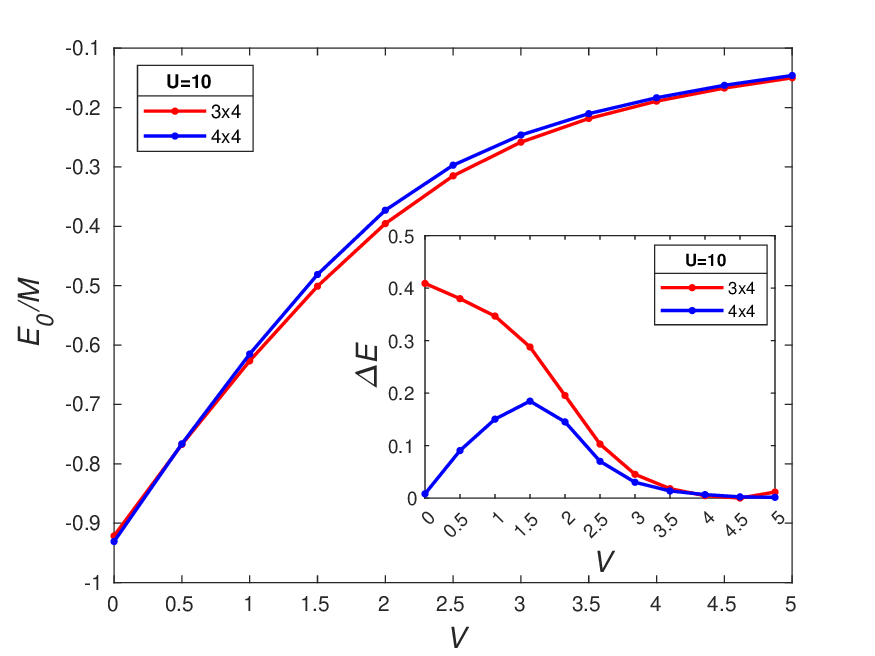}
\caption{Ground-state energy per site $E_0/M$ versus NN repulsion $V$ for quarter-filled extended Hubbard clusters at $U=10$. The inset shows the excitation gap $\Delta E$. Both clusters become energetically flatter at large $V$; the gap collapses beyond $V\!\gtrsim\!3$, indicating near-degenerate charge configurations.}
\label{fig:E0gapehmu10}
\end{figure}

Figure~\ref{fig:E0gapehmu10} shows the ground-state energy per site $E_0/M$ and the excitation gap $\Delta E=E_1-E_0$ (inset) as functions of $V$. In both clusters, $E_0/M$ increases monotonically with $V$, reflecting the additional Coulomb energy cost of adjacent occupancy. The gaps, however, exhibit distinct behaviors: for the $3\times4$ lattice, $\Delta E$ decreases smoothly from $\sim0.41$ at $V=0$ to nearly zero at $V=5$, while for the $4\times4$ lattice it first rises slightly, peaks near $V\!\approx\!1.5$, and then collapses toward zero beyond $V\!\gtrsim\!3$. The simultaneous flattening of $E_0/M$ and collapse of $\Delta E$ signal a suppression of charge excitations and a crossover to a correlated, nearly degenerate manifold of charge-ordered states. At large $V$, the nearest-neighbor repulsion forces electrons to avoid adjacent sites, creating multiple low-energy charge configurations that are nearly degenerate --- hence the gap collapses and the energy flattens \cite{Aaron2014}.
Finite-size scaling behavior of the excitation gap is discussed in Appendix \ref{FSSU10}.

\subsubsection{Local observables: double occupancy and local moment}

\begin{figure}[t]
\centering
\includegraphics[width=0.48\textwidth]{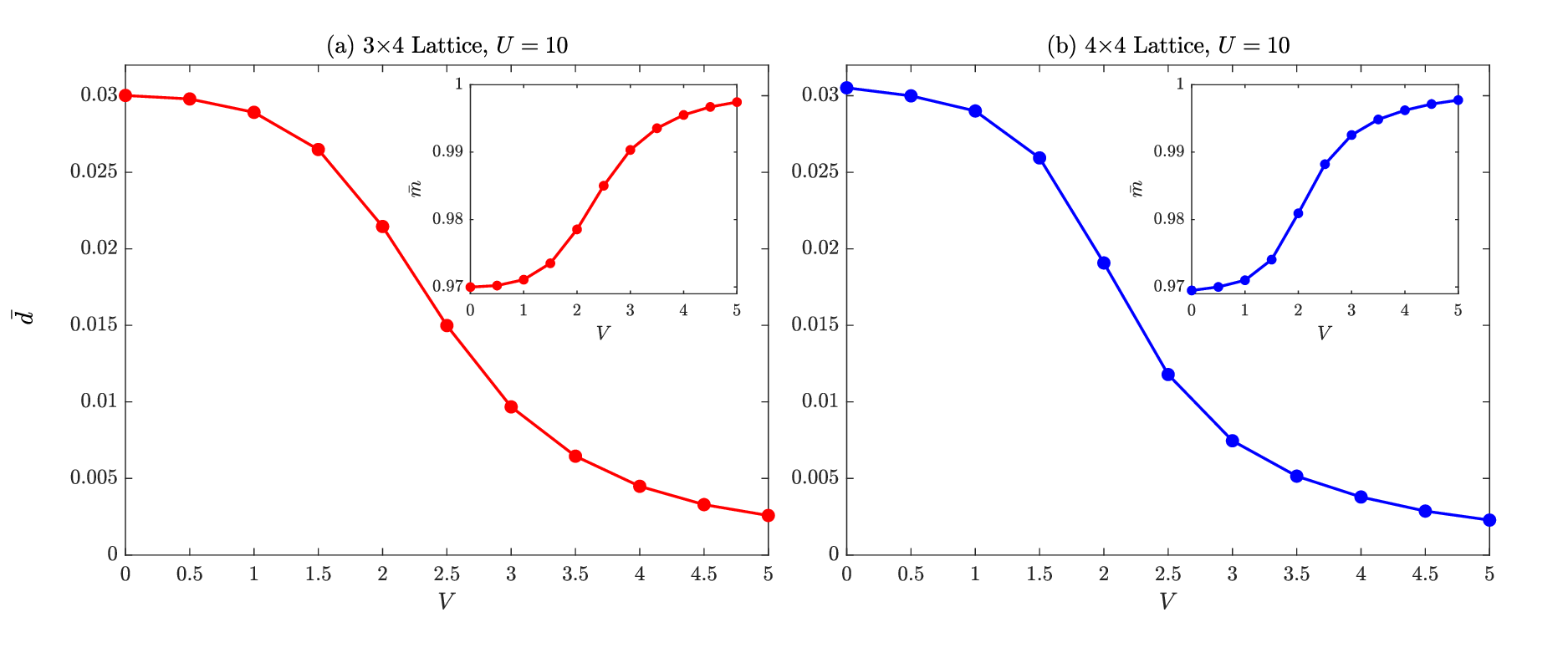}
\caption{Average double occupancy $\bar d$ and local moment $\bar m$ (inset) as functions of nearest-neighbor repulsion $V$ at fixed $U=10$. Both clusters show suppression of $\bar d$ and growth of $\bar m$, signifying enhanced localization and local-moment stabilization.}
\label{fig:dbclcmu10}
\end{figure}
The suppression of charge fluctuations is further evident in Fig.~\ref{fig:dbclcmu10}, which shows the average double occupancy $\bar d$ and local moment $\bar m$. At $U=10$, $\bar d$ is already small ($\approx0.03$) and decreases steadily with $V$, while $\bar m$ increases correspondingly, approaching full local-moment formation. These opposing trends confirm that the large on-site repulsion freezes double occupancy, and the added $V$ further enhances charge separation. The $4\times4$ cluster consistently exhibits smaller $\bar d$ and larger $\bar m$,
consistent with stronger localization and better development of short-range AFM correlations.
Finite-size scaling analysis of the average double occupancy $\bar d$ is presented in Appendix \ref{FSSU10}.

\subsubsection{Charge and spin structure factors}

\begin{figure}[t]
\centering
\includegraphics[width=0.50\textwidth]{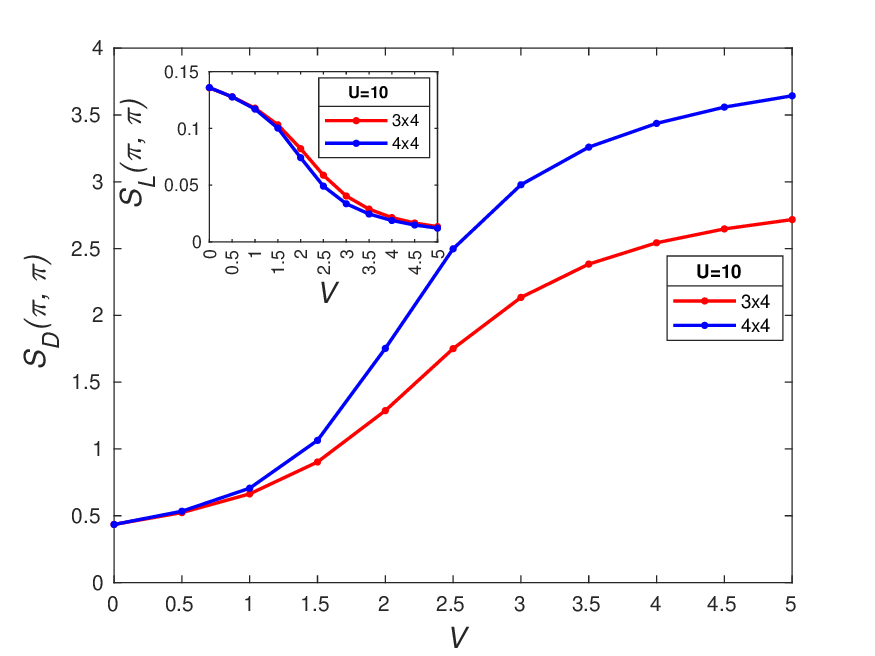}
\caption{Charge and spin structure factors $S_D(\pi,\pi)$ and $S_L(\pi,\pi)$ versus $V$ for the extended Hubbard model at $U=10$. Increasing $V$ strengthens charge ordering while suppressing AFM spin correlations, demonstrating a smooth evolution from a Mott-like to a CDW-dominated regime.}
\label{fig:csfssfu10}
\end{figure}

The interplay between charge and spin ordering is summarized in Fig.~\ref{fig:csfssfu10}. The charge structure factor $S_D(\pi,\pi)$ increases sharply with $V$, revealing the progressive establishment of charge-density-wave (CDW) order. In contrast, the spin structure factor $S_L(\pi,\pi)$ (inset) decreases steadily, showing that long-range AFM correlations are suppressed once charge alternation becomes energetically favorable. The crossover between these two trends occurs near $V\!\approx\!1.5$, marking the onset of CDW dominance. The larger $4\times4$ cluster exhibits a steeper $S_D(\pi,\pi)$ rise and a stronger suppression of $S_L(\pi,\pi)$, consistent with its enhanced capacity for charge modulation.

\subsubsection{Local-singlet pairing correlations}

\begin{figure}[t]
\centering
\includegraphics[width=0.50\textwidth]{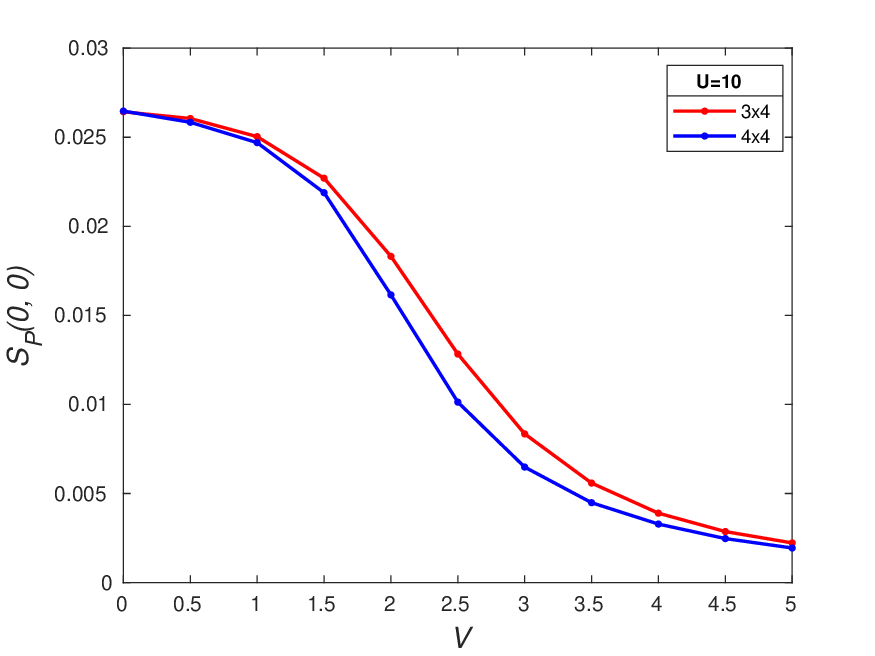}
\caption{Local-singlet pairing structure factor $S_P(0,0)$ versus $V$ at fixed $U=10$. Pairing correlations diminish steadily as intersite repulsion strengthens, consistent with the dominance of charge ordering at large $V$.}
\label{fig:psfu10}
\end{figure}

The evolution of local-singlet pairing correlations, quantified by $S_P(0,0)$, is shown in Fig. \ref{fig:psfu10}. At small $V$, a finite value of $S_P(0,0)$ persists due to residual virtual charge fluctuations that allow transient double occupancy even at large $U$. As $V$ increases, $S_P(0,0)$ decreases monotonically, indicating that intersite repulsion suppresses the charge fluctuations and inhibits local pairing correlations. This decline becomes sharp beyond $V\!\gtrsim\!2$, coinciding with the rapid growth of $S_D(\pi,\pi)$ and confirming that CDW order dominates the strong-coupling regime.

\subsubsection{PCA of charge, spin and pairing correlations}

To provide a unified statistical perspective, we again apply PCA to the site-resolved charge, spin and local-pairing correlation matrices for both clusters.

\begin{figure}[t]
\centering
\includegraphics[width=0.48\textwidth]{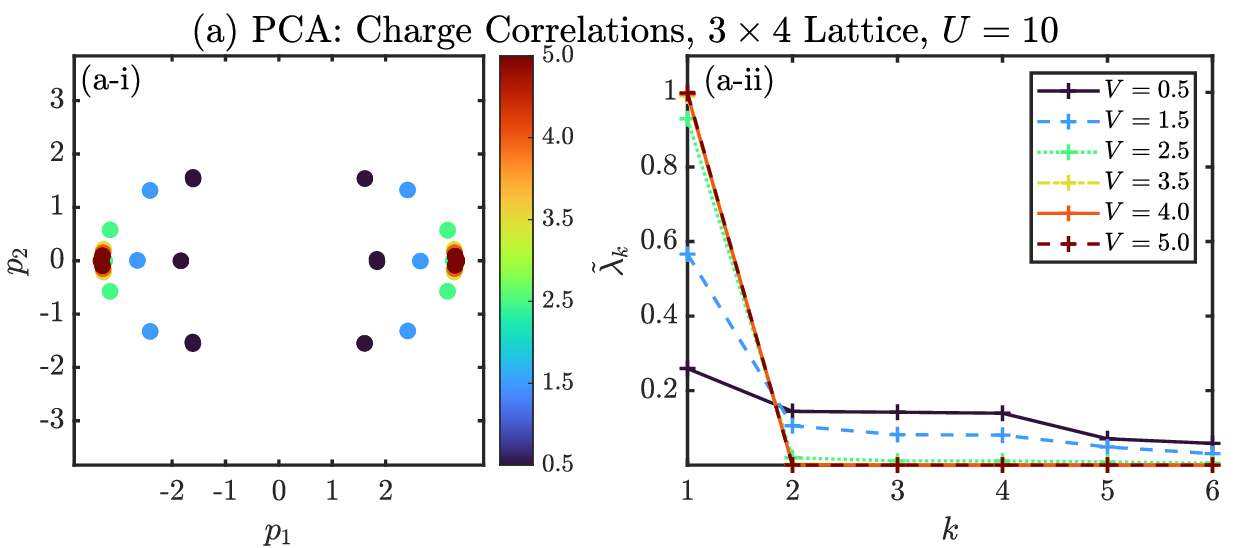}
\hspace{0.2cm}
\includegraphics[width=0.48\textwidth]{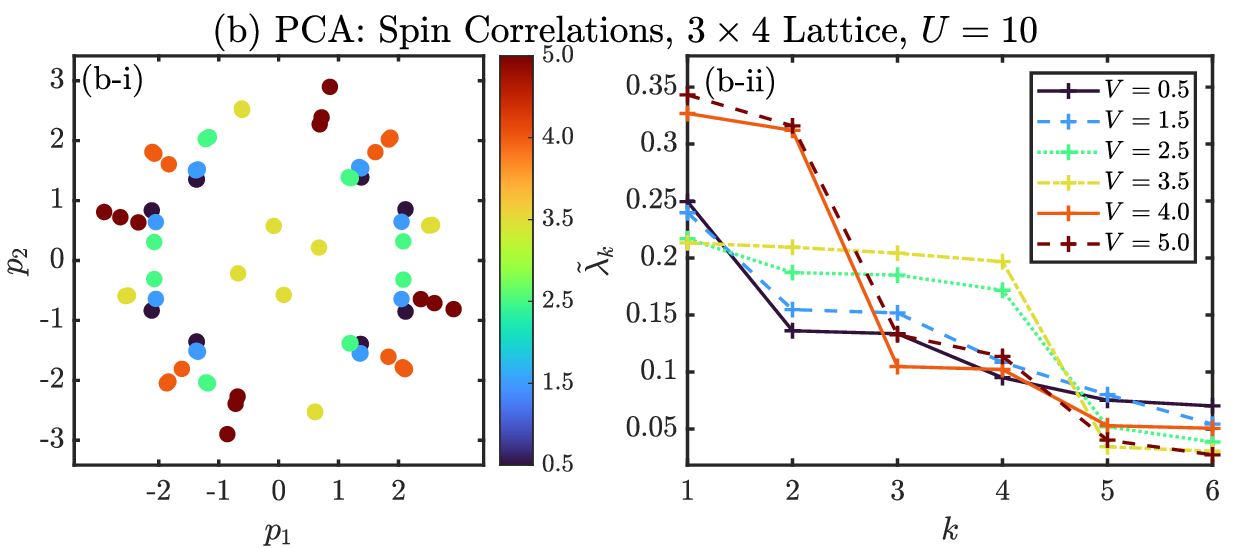}
\hspace{0.2cm}
\includegraphics[width=0.48\textwidth]{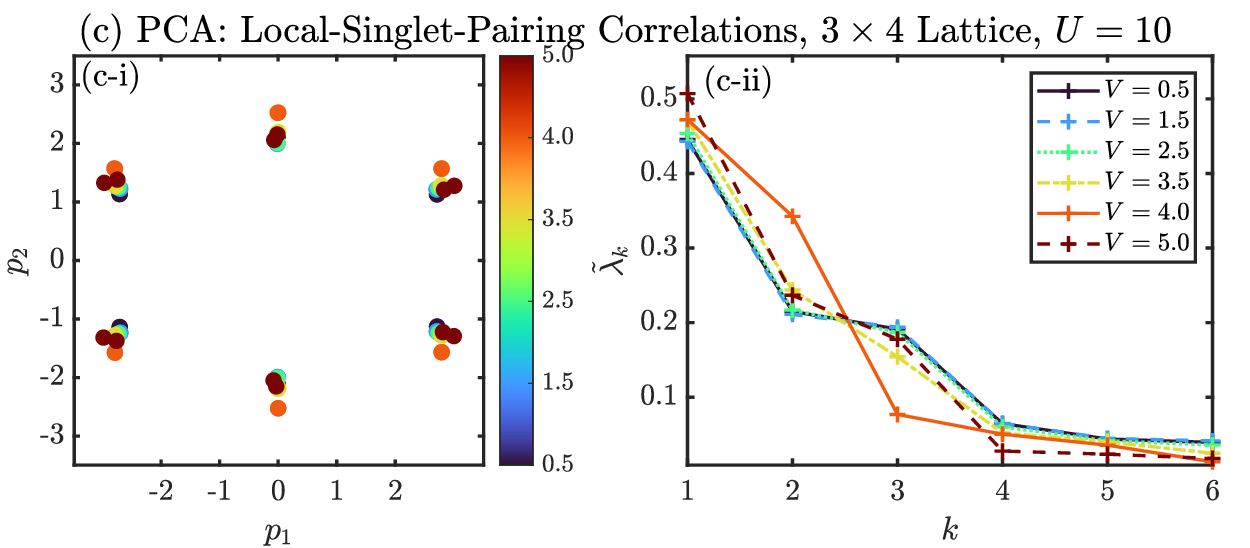}
\caption{Principal-component analysis (PCA) of (a) charge, (b) spin, and (c) local-singlet-pairing correlation matrices for the $3\times4$ cluster at $U=10$ as $V$ increases. Beyond $V\!\approx\!2$, a single charge component dominates ($\tilde{\lambda}_1\!\to\!1$), identifying the transition to a charge-ordered regime, while spin and pairing channels remain multi-modal.}
\label{fig:3x4pcau10}
\end{figure}

Figure \ref{fig:3x4pcau10} shows the $3\times4$ results.
In the charge sector [Fig. \ref{fig:3x4pcau10}(a)], projections onto the first two principal components remain scattered at small $V$, with the variance distributed over several components ($\tilde{\lambda}_1\!\approx\!0.26$, $\tilde{\lambda}_2$-$\tilde{\lambda}_4\!\approx\!0.14$), indicating charge fluctuations where no single spatial pattern dominates, but their superposition forms an emergent spatial distribution. 
Analysis of higher components ($p_3$, $p_4$) shows no additional distinct structure beyond that captured by $p_1$ and $p_2$, confirming that the observed multi-component character reflects distributed variance rather than hidden ordered patterns.
As $V$ increases beyond $\sim 2$, the projections reorganize into clearly separated clusters and the leading eigenvalue rises sharply ($\tilde{\lambda}_1\!\to\!0.9$-$1.0$). This strong concentration of variance into a single dominant principal component signals the emergence of CDW-dominated phase, in which charge correlations become locked into one coherent spatial pattern with minimal mixing from competing subleading principal components.
In the spin sector [Fig.~\ref{fig:3x4pcau10}(b)], projections onto the first two principal components ($p_1$, $p_2$) remain distributed across several loosely grouped clusters with only weak dependence on $V$, and the explained-variance spectrum retains a multi-component character with $\tilde{\lambda}_1$ and $\tilde{\lambda}_2$ of comparable weight. This indicates that spin correlations remain distributed across multiple principal components, without condensation into a single dominant principal component.
In the pairing sector [Fig.~\ref{fig:3x4pcau10}(c)], the projections onto ($p_1$, $p_2$) form clusters of overlapping distributions whose arrangement exhibits minimal variation with $V$, while two leading principal components remain significant ($\tilde{\lambda}_1 \approx 0.45$, $\tilde{\lambda}_2 \approx 0.20$). This indicates that pairing correlations are distributed across multiple components embedded within the charge-ordered background.

\begin{figure}[t]
\centering
\includegraphics[width=0.48\textwidth]{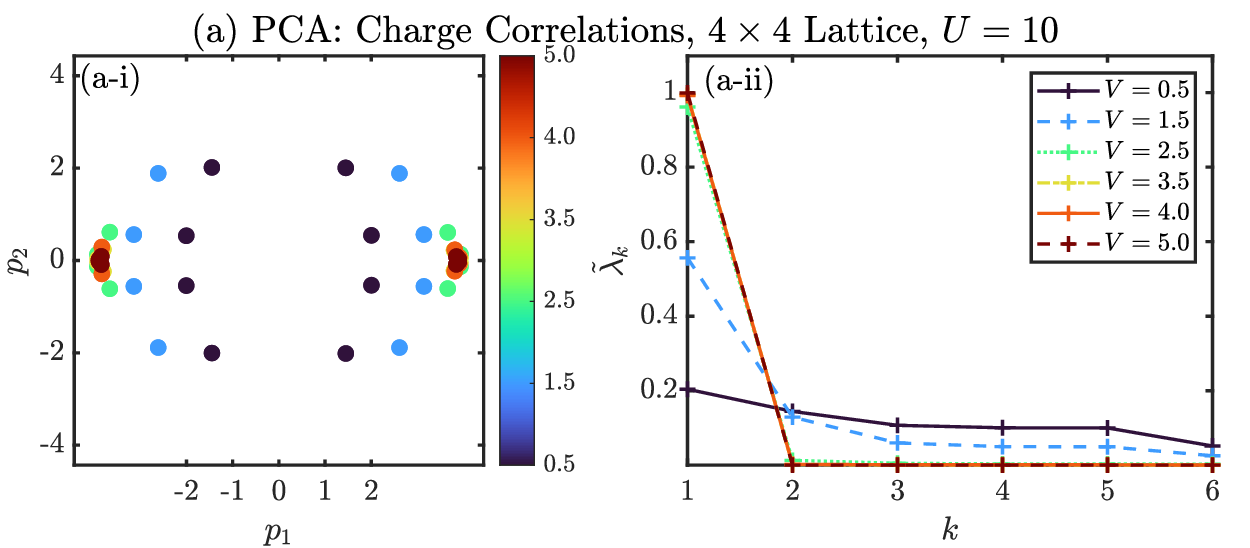}
\hspace{0.2cm}
\includegraphics[width=0.48\textwidth]{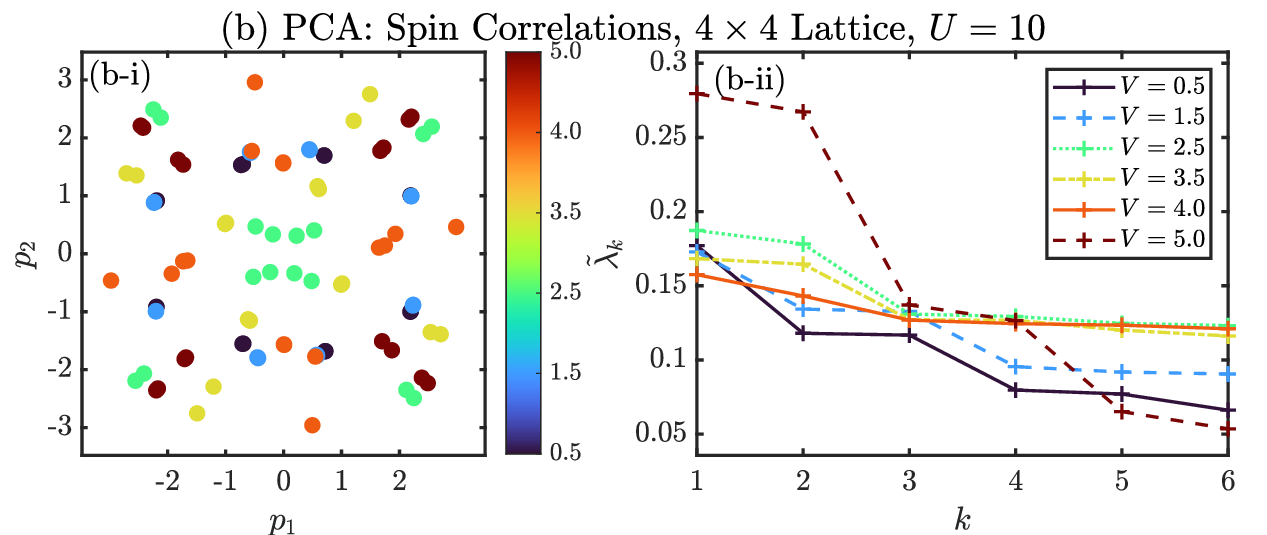}
\hspace{0.2cm}
\includegraphics[width=0.48\textwidth]{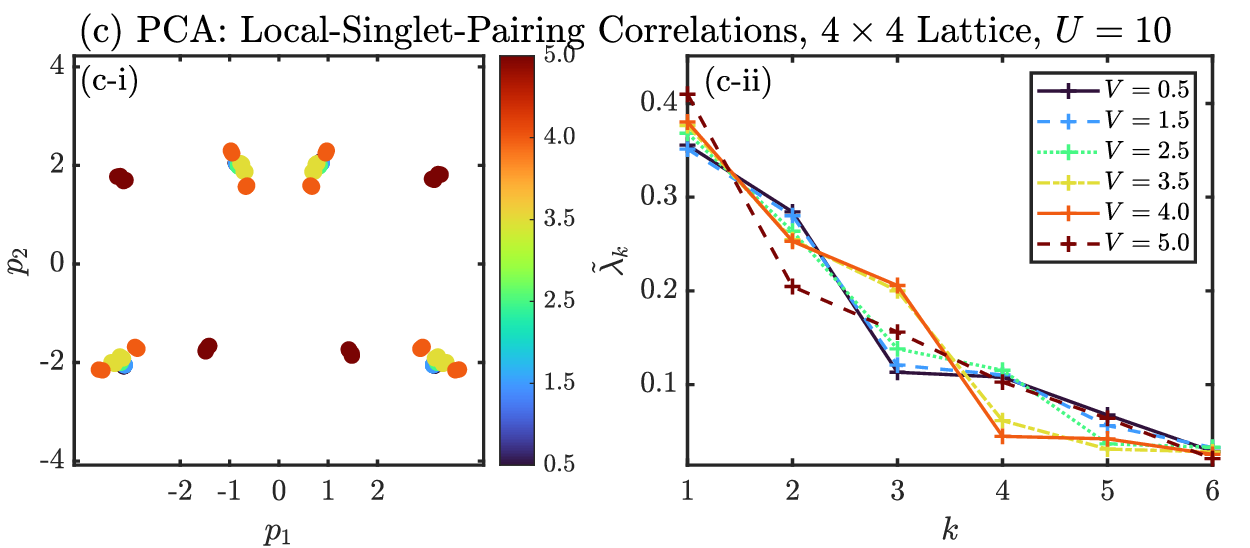}
\caption{Principal-component analysis (PCA) of (a) charge, (b) spin, and (c) local-singlet-pairing correlations for the $4\times4$ cluster at $U=10$. The variance condenses into a single charge mode at large $V$ ($\tilde{\lambda}_1\!\approx\!1$), while spin and pairing sectors remain multi-component, indicating dominance of charge-density ordering.}
\label{fig:4x4pcau10}
\end{figure}

The $4\times 4$ cluster exhibits a similar but sharper evolution (Fig.\ref{fig:4x4pcau10}). In the charge sector [Fig.\ref{fig:4x4pcau10}(a)] at weak $V$, several leading principal components coexist, but beyond $V\!\gtrsim\!2.5$ the first component overwhelmingly dominates ($\tilde{\lambda}_1\!\to\!1$), signaling complete condensation into a single CDW-dominated principal component. Spin correlations remain fragmented and weak ($\tilde{\lambda}_1$-$\tilde{\lambda}_3\!\approx\!0.15$), and pairing retains two near-degenerate components ($\tilde{\lambda}_1,\tilde{\lambda}_2\!\approx\!0.35$-$0.40$), implying residual two-channel singlet fluctuations. However, because the $\tilde{\lambda}_k$ values are relative within each sector, direct cross‑channel comparison is not meaningful; the hierarchy is instead established by combining the PCA spectra with complementary observables such as the flattening of $E_0/M$, collapse of $\Delta E$ and suppression of $\bar{d}$. Thus, PCA reveals a clear hierarchy: charge order dominates, spin order destroys, and pairing survives only locally.

\begin{figure*}[t]
\centering
\includegraphics[width=2.0\columnwidth]{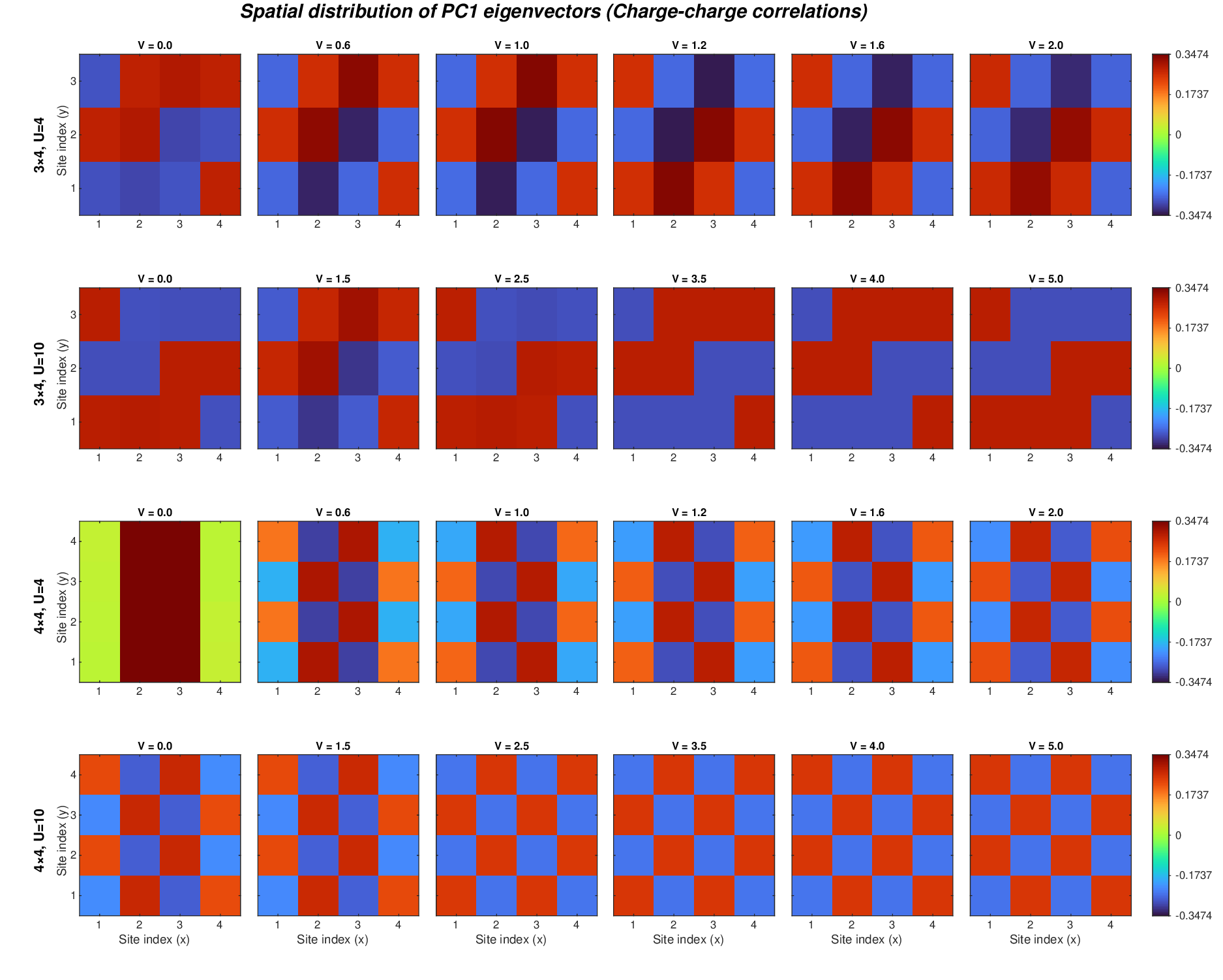}
\caption{Representative eigenvectors $w_1$ of the charge correlation matrix for $3\times4$ and $4\times4$ clusters at $U=4$ and $U=10$, for both simple ($V=0$) and extended ($V>0$) Hubbard models. Each panel shows the real-space pattern of $w_1$ across the lattice sites.}
\label{fig:combined_charge_eigvecs}
\end{figure*}

\subsubsection{Summary of the strong-coupling regime}

In the strong-coupling regime ($U=10$), the PCA analysis reveals a clear hierarchy among the competing fluctuation channels.  
The charge sector undergoes a pronounced dimensional reduction of its variance spectrum, with the leading principal component $\tilde{\lambda}_1$ saturating close to unity as $V$ increases, signifying the emergence of a robust charge-ordered phase.  
In contrast, spin fluctuations fragment into several low-weight principal components, consistent with the suppression of antiferromagnetic order by the alternating charge pattern.  
Local-pairing correlations persist only as weak, two-component fluctuations embedded within this charge-ordered background.  
Compared with the intermediate-coupling case ($U=4$), the crossover to dominant charge order now occurs at smaller $V/U$, reflecting enhanced electron localization and reduced kinetic mobility in the strong-coupling limit.  
Overall, the PCA framework provides a transparent and an unsupervised picture of how competing correlations reorganize across interaction scales in finite Hubbard clusters.

\subsection*{Unified Physical Interpretation of the Leading PCA Eigenvectors}
To elucidate the physical meaning of the leading principal components, we jointly analyze the leading eigenvectors $w_k$ obtained from the charge-correlation matrices  of both the simple and extended Hubbard models at quarter filling. While the explained-variance ratios $\tilde{\lambda}_k$ quantify the relative strength of each fluctuation channel, the eigenvectors reveal their real-space \emph{form factors}. Mapping $w_k$ onto the lattice identifies the dominant spatial organization of charge correlations at different interaction strengths and cluster sizes. Although only the charge-sector eigenvectors are plotted here, analogous behavior is verified for the spin and pairing correlation matrices (\emph{results not shown}).

Across all six representative cases (Fig.~\ref{fig:combined_charge_eigvecs}),
a common pattern emerges:

\textbf{(i) Simple Hubbard model (\(V=0\)):}  
For the $3\times 4$ cluster at both $U=4$ and $U=10$, the leading eigenvector $w_1$ exhibits a clear checkerboard pattern, consistent with a $(\pi,\pi)$ CDW form factor. On the $4\times 4$ cluster, $w_1$ shows a stripe-like modulation at $U=4$ but evolves to checkerboard at $U=10$, highlighting finite-size sensitivity. In all cases, the staggered character strengthens with $U$, reflecting progressive local-moment formation. Appendix \ref{subpcaeigvec} presents representative subleading principal component eigenvectors.

\textbf{(ii) Extended Hubbard model (\(V>0\)):}  
Introducing $V>0$ reinforces and sharpens the staggered charge modulation already present at $V=0$. For both cluster sizes and both values of $U$ (=4,10), the leading eigenvector $w_1$ retains its checkerboard-like form factor, while the explained-variance ratio $\tilde \lambda_k$ increases markedly with $V$. At moderate $V$, where the variance spectrum remains partially mixed, subleading eigenvectors (as shown in Appendix \ref{subpcaeigvec}) exhibit shorter-range or irregular modulations that do not qualitatively alter the overall staggered character of the dominant pattern. As $V$ grows (particularly beyond $V\approx 2-2.5$), higher components ($w_2$, $w_3$) lose weight rapidly, and $w_1$ captures nearly all the variance. The spatial form of $w_1$ undergoes only minor quantitative adjustments, confirming that intersite repulsion primarily concentrates fluctuations into the pre-existing staggered CDW channel rather than inducing a new symmetry.

\textbf{(iii) Parallel trends in spin and pairing sectors (not shown):}  
Although only the charge-sector eigenvectors are plotted, analogous behavior is verified for
the spin and pairing correlation matrices.  
For spins, the leading eigenvector $w_1^S$ acquires a staggered $(\pi,\pi)$ form as $U$
increases, consistent with emerging short-range antiferromagnetism.
In the pairing sector, two nearly degenerate principal components persist, reflecting
competing local singlet correlations rather than a single dominant superconducting form factor.
These results confirm that PCA consistently extracts the expected real-space motifs across all
interaction channels.

\textbf{(iv) Unified interpretation:}  
Taken together, the eigenvector analyses across the six cases show a continuous evolution
from delocalized metallic correlations to localized charge order.
The PCA automatically identifies the symmetry and spatial pattern of the leading principal component---checkerboard for charge, staggered for spin, thereby bridging statistical variance with microscopic order.  
This unified treatment avoids repetition while emphasizing that both the simple and extended
models encode the same underlying physics: the progressive condensation of correlation
variance into well-defined form factors as interaction strengths grow.

\section{\label{summry}Summary and Conclusions}
The PCA eigenvector analysis provides a compact view of how competing fluctuation channels reorganize with increasing $U$ and $V$. We now discuss the broader implications of these trends in relation to correlated electron physics. In summary, we have applied PCA to exact-diagonalization data of quarter-filled Hubbard clusters, identifying the dominant fluctuation channels and their real-space form factors.
While these models have been extensively explored through analytical and numerical techniques,
our approach combines exact cluster spectra with
\emph{principal component analysis} (PCA) of correlation matrices,
providing a compact, model-agnostic framework
for identifying dominant fluctuation channels and crossover behavior
without imposing predefined order parameters.

For the simple Hubbard model ($V=0$), increasing on-site repulsion $U$ drives a smooth crossover from a delocalized metallic-like regime to a spin-correlated Mott-like state. The PCA analysis quantitatively visualizes this crossover: the leading variance shifts from charge to spin sectors, and the clustering patterns in the principal-component space evolve from diffuse charge-dominated clouds to compact spin-correlated structures. These features provide an unbiased statistical fingerprint of local-moment formation and the onset of short-range antiferromagnetic correlations, consistent with the evolution of the structure factors $S_D(\pi,\pi)$ and $S_L(\pi,\pi)$.
These conclusions are further examined by a finite-size scaling analysis (Appendix \ref{FSS}), which confirms that the observed gap suppression and crossover features are smooth finite-size effects.

Introducing the nearest-neighbor repulsion $V$ enriches the correlation landscape.
At intermediate coupling ($U=4$),
the system exhibits a delicate balance between local-moment formation and charge modulation.
With increasing $V$, the charge structure factor $S_D(\pi,\pi)$ grows rapidly,
and PCA reveals a progressive condensation of variance into a single dominant principal component, signifying the emergence of charge-density-wave (CDW) ordering.
Spin correlations weaken, while local-singlet pairing shows a mild enhancement at moderate $V$
followed by suppression at larger $V$, indicating a transient coexistence regime
of fluctuating pairing and charge order in small clusters.

In the strong-coupling regime ($U=10$),
the interplay between $U$ and $V$ yields a hierarchy of correlation scales.
Double occupancy becomes strongly suppressed and local moments saturate,
while the nearest-neighbor repulsion reorganizes charge degrees of freedom into a robust CDW pattern.
The PCA results capture this transformation in a single glance:
the leading eigenvalue $\tilde{\lambda}_1$ in the charge sector
rises sharply to unity as $V$ increases, while the spin and pairing spectra fragment into
weak, multi-component fluctuations.
The dimensional reduction of the variance spectrum thus acts as a quantitative marker
of emergent charge order and suppressed magnetic coherence.
Compared with the $U=4$ case, the crossover to charge-dominated behavior
occurs at a smaller $V/U$ ratio, underscoring the enhanced localization at large $U$.

Overall, our findings establish PCA as a powerful complementary tool
for analyzing strongly correlated systems, even in small clusters
where traditional order parameters or finite-size scaling may fail to reveal clear transitions.
By treating correlation matrices as high-dimensional observables,
PCA identifies the dominant fluctuation subspace and traces its evolution
across interaction parameters in an unbiased way.
This approach offers a unified framework for visualizing the competition
among charge, spin, and pairing orders within finite Hubbard systems,
and can be readily extended to larger clusters,
multi-orbital models, or dynamical datasets from DMFT and quantum Monte Carlo simulations.

\medskip
\noindent

The present framework opens a route to a more systematic integration
of machine-learning diagnostics with exact and dynamical methods in correlated-electron physics.
Future work may combine PCA with nonlinear manifold techniques (e.g., t-SNE, UMAP)
to capture subtle order-parameter topology,
or apply time-dependent PCA to trace real-time reorganization of correlations
in driven Hubbard clusters.
Such data-driven analyses promise to complement conventional observables
and to deepen our microscopic understanding of quantum correlations
in low-dimensional materials.

\section*{Acknowledgements}

MFE and MAHA acknowledge the support from National PARAM Supercomputing Facility (NPSF), C-DAC, Pune, India, for providing computing facility for part of this work.

\begin{widetext}
\section*{Appendix}
\appendix

\section{Eigenvalues and corresponding eigenvectors for the non-interacting ($U,V=0$) case of $3\times 4$ cylindrical cluster.}
\label{3x4mopbc}
We construct the single-particle MO states for $3\times 4$ cylindrical cluster with PBC along the x-direction and OBC along the y-direction, capturing the lattices periodic and finite structure. For the x-direction with PBC, the wavefunction is a Bloch state, $\psi_n(x)=e^{ik_x x}u_{k_x}(x)$, where the wave vectors are $k_x=\frac{2n\pi}{4}$ $(n=0,1,2,3)$ and $u_{k_x}(x)$ is periodic function with lattice spacing $a=1$. In the tight-binding approximation of the Hubbard model, the periodic function $u_{k_x}(x)$ is constant (\textit{i.e.,} $u_{k_x}(x)=1$) due to the single orbital per site and the discrete lattice, simplifying the x-component to $\psi_n(x)=\frac{1}{\sqrt{4}} e^{ik_x x}$. For the y-direction with OBC, we introduce virtual sites at $y=0$ and $y=4$ to enforce zero wavefunction amplitude at the boundaries, yielding a standing wave, $\psi_m(y)=\sqrt{\frac{2}{4}}sin(k_y y)$, with wave vectors $k_y=\frac{m\pi}{4}$ $(m=1,2,3)$, consistent with the fixed boundary conditions. The MO wavefunction is the product $\phi_{k_x,k_y}(x,y)=\frac{1}{2\sqrt{2}}e^{i\frac{2n\pi}{4}x}sin(\frac{m\pi y}{4})$, forming the state     
\begin{equation}
|k_x, k_y, \sigma\rangle=\sum_{x=1}^4\sum_{y=1}^3\phi_{k_x,k_y}(x,y)c_{x,y,\sigma}^\dagger|0\rangle,
\end{equation}
where $x$ labels the sites along the x-direction (1 to 4) and $y$ labels the rows (1 to 3). The dispersion relation is given by  
\begin{equation}
\epsilon_{n,m}=-2tcos(\frac{2n\pi}{4})-2tcos(\frac{m\pi}{4}), 
\end{equation}
yielding 12 distinct energy levels (molecular orbitals) as listed below. 

\begin{equation}
|k_x=0, k_y=\frac{\pi}{4},\lambda = -2-\sqrt{2} \rangle = -\frac{1}{4}(|1\rangle-|2\rangle+|3\rangle-|4\rangle+|9\rangle-|10\rangle+|11\rangle-|12\rangle)+\frac{1}{2\sqrt{2}}(|5\rangle-|6\rangle+|7\rangle-|8\rangle) \\
\end{equation}
\begin{equation}
|k_x=0, k_y=\frac{\pi}{2}, \lambda = -2 \rangle = \frac{1}{2\sqrt{2}}(|1\rangle-|2\rangle+|3\rangle-|4\rangle-|9\rangle+|10\rangle-|11\rangle+|12\rangle)+0(|5\rangle+|6\rangle+|7\rangle+|8\rangle) \\
\end{equation}
\begin{equation}
|k_x=\frac{\pi}{2}, k_y=\frac{\pi}{4}, \lambda = -\sqrt{2};1 \rangle = \frac{1}{2\sqrt{2}}(|1\rangle-|3\rangle+|9\rangle-|11\rangle)+0(|2\rangle+|4\rangle+|6\rangle+|8\rangle+|10\rangle+|12\rangle) - \frac{1}{2}(|5\rangle-|7\rangle) \\
\end{equation}
\begin{equation}
|k_x=\frac{3\pi}{2}, k_y=\frac{\pi}{4}, \lambda = -\sqrt{2};2 \rangle = 0(|1\rangle+|3\rangle+|5\rangle+|7\rangle+|9\rangle+|11\rangle)+\frac{1}{2\sqrt{2}}(|2\rangle-|4\rangle+|10\rangle-|12\rangle)-\frac{1}{2}(|6\rangle-|8\rangle) \\
\end{equation}
\begin{equation}
|k_x=0, k_y=\frac{3\pi}{4}, \lambda = -2+\sqrt{2} \rangle = \frac{1}{4}(|1\rangle-|2\rangle+|3\rangle-|4\rangle+|9\rangle-|10\rangle+|11\rangle-|12\rangle)+\frac{1}{2\sqrt{2}}(|5\rangle-|6\rangle+|7\rangle-|8\rangle) \\
\end{equation}
\begin{equation}
|k_x=\frac{\pi}{2}, k_y=\frac{\pi}{2}, \lambda = 0;1 \rangle = 0(|1\rangle+|3\rangle+|5\rangle+|6\rangle+|7\rangle+|8\rangle+|9\rangle+|11\rangle)+\frac{1}{2}(|2\rangle-|4\rangle-|10\rangle+|12\rangle) \\
\end{equation}
\begin{equation}
|k_x=\frac{3\pi}{2}, k_y=\frac{\pi}{2}, \lambda = 0;2 \rangle = \frac{1}{2}(|1\rangle-|3\rangle-|9\rangle+|11\rangle)+0(|2\rangle+|4\rangle+|5\rangle+|6\rangle+|7\rangle+|8\rangle+|10\rangle+|12\rangle) \\
\end{equation}
\begin{equation}
|k_x=\pi, k_y=\frac{\pi}{4}, \lambda = 2-\sqrt{2} \rangle = -\frac{1}{4}(|1\rangle+|2\rangle+|3\rangle+|4\rangle+|9\rangle+|10\rangle+|11\rangle+|12\rangle)+\frac{1}{2\sqrt{2}}(|5\rangle+|6\rangle+|7\rangle+|8\rangle) \\
\end{equation}
\begin{equation}
|k_x=\frac{\pi}{2}, k_y=\frac{3\pi}{4}, \lambda = \sqrt{2};1 \rangle = \frac{1}{2\sqrt{2}}(|1\rangle-|3\rangle+|9\rangle-|11\rangle)+0(|2\rangle+|4\rangle+|6\rangle+|8\rangle+|10\rangle+|12\rangle)+\frac{1}{2}(|5\rangle-|7\rangle) \\
\end{equation}
\begin{equation}
|k_x=\frac{3\pi}{2}, k_y=\frac{3\pi}{4}, \lambda = \sqrt{2};2 \rangle = 0(|1\rangle+|3\rangle+|5\rangle+|7\rangle+|9\rangle+|11\rangle)-\frac{1}{2\sqrt{2}}(|2\rangle-|4\rangle+|10\rangle-|12\rangle)-\frac{1}{2}(|6\rangle-|8\rangle) \\
\end{equation}
\begin{equation}
|k_x=\pi, k_y=\frac{\pi}{2}, \lambda = 2 \rangle = -\frac{1}{2\sqrt{2}}(|1\rangle+|2\rangle+|3\rangle+|4\rangle-|9\rangle-|10\rangle-|11\rangle-|12\rangle)+0(|5\rangle+|6\rangle+|7\rangle+|8\rangle) \\
\end{equation}
\begin{equation}
|k_x=\pi, k_y=\frac{3\pi}{4}, \lambda = 2+\sqrt{2} \rangle = \frac{1}{4}(|1\rangle+|2\rangle+|3\rangle+|4\rangle+|9\rangle+|10\rangle+|11\rangle+|12\rangle)+\frac{1}{2\sqrt{2}}(|5\rangle+|6\rangle+|7\rangle+|8\rangle)  \\
\end{equation}

\section{Eigenvalues and corresponding eigenvectors for the non-interacting ($U,V=0$) case of $4\times 4$ cylindrical cluster.}
\label{4x4mopbc}
The single-particle MO states for $4\times 4$ cylindrical cluster results from the combination of plane wave due to PBC in the x-direction and standing wave due to OBC in the y-direction. For PBC in the x-direction the wavefunction is a plane wave written as
\begin{equation}
\psi_{k_x}(x)=\frac{1}{\sqrt{N_x}}e^{ik_x x},
\end{equation}   
where $k_x=\frac{2\pi n_x}{N_x}$, $n_x=0,1,2,3$ and $N_x=4$. Therefore, the allowed momenta in the x-direction are $k_x=0, \pi/2, \pi, 3\pi/2(-\pi/2)$. In the y-direction, OBC leads to standing waves. The eigenstates for a 1D chain of length $N_y=4$ is written as 
\begin{equation}
\psi_{n_y}(y)=\sqrt{\frac{2}{N_y+1}}sin(\frac{n_y\pi y}{N_y+1})=\sqrt{\frac{2}{5}}sin(\frac{n_y\pi y}{5}),
\end{equation}
where $n_y=1,2,3,4$. The resulting MO wavefunction is 
\begin{equation}
\phi_{k_x, n_y}(x,y)=\frac{1}{\sqrt{4}}e^{ik_x x}\sqrt{\frac{2}{5}}sin(\frac{n_y\pi y}{5}). 
\end{equation} 
The total energy for a state ($k_x, n_y$) is the sum of contributions from x- and y-directions given by
\begin{equation}
\epsilon_{k_x, n_y}=-2tcos(k_x)-2tcos(\frac{n_y\pi}{5}), 
\end{equation}
yielding 16 distinct MO states. Each state is labeled by ($k_x, n_y$), and the corresponding eigenvectors are the products of plane waves in x-direction  and sine waves in y-direction and is listed below.
\begin{eqnarray}
|k_x=0, n_y=1; \lambda = -3.618 \rangle &=& -0.1859(|1\rangle-|2\rangle +|3\rangle-|4\rangle-|13\rangle+|14\rangle-|15\rangle+|16\rangle)+0.3008(|5\rangle-|6\rangle  \nonumber \\
&+&|7\rangle-|8\rangle-|9\rangle+|10\rangle-|11\rangle+|12\rangle)
\end{eqnarray}
\begin{eqnarray}
|k_x=0, n_y=2; \lambda = -2.618 \rangle &=& -0.3008(|1\rangle-|2\rangle +|3\rangle-|4\rangle+|13\rangle-|14\rangle+|15\rangle-|16\rangle)+0.1859(|5\rangle-|6\rangle \nonumber \\
&+&|7\rangle-|8\rangle+|9\rangle-|10\rangle+|11\rangle-|12\rangle)
\end{eqnarray}
\begin{eqnarray}
|k_x=\pi/2, n_y=1; \lambda = -1.618;1 \rangle &=& 0(|1\rangle+|3\rangle +|5\rangle+|7\rangle+|9\rangle+|11\rangle+|13\rangle+|15\rangle)-0.2629(|2\rangle-|4\rangle \nonumber \\
&-&|14\rangle+|16\rangle)+0.4253(|6\rangle-|8\rangle-|10\rangle+|12\rangle)
\end{eqnarray}
\begin{eqnarray}
|k_x=3\pi/2, n_y=1; \lambda = -1.618;2 \rangle &=& 0.2629(|1\rangle-|3\rangle -|13\rangle+|15\rangle)+ 0(|2\rangle+|4\rangle+|6\rangle+|8\rangle+|10\rangle \nonumber \\
&+&|12\rangle+|14\rangle+|16\rangle)-0.4253(|5\rangle-|7\rangle-|9\rangle+|11\rangle)
\end{eqnarray}
\begin{eqnarray}
|k_x=0, n_y=3; \lambda = -1.382 \rangle &=& 0.3008(|1\rangle-|2\rangle +|3\rangle-|4\rangle-|13\rangle+|14\rangle-|15\rangle+|16\rangle)+0.1859(|5\rangle-|6\rangle  \nonumber \\ 
&+&|7\rangle-|8\rangle-|9\rangle+|10\rangle-|11\rangle+|12\rangle)
\end{eqnarray}
\begin{eqnarray}
|k_x=\pi/2, n_y=2; \lambda = -0.618;1 \rangle &=& 0(|1\rangle+|3\rangle +|5\rangle+|7\rangle+|9\rangle+|11\rangle+|13\rangle+|15\rangle)+0.4253(|2\rangle-|4\rangle \nonumber \\
&+&|14\rangle-|16\rangle)- 0.2629(|6\rangle-|8\rangle+|10\rangle-|12\rangle)
\end{eqnarray}
\begin{eqnarray}
|k_x=3\pi/2, n_y=2; \lambda = -0.618;2 \rangle &=& 0.4253(|1\rangle-|3\rangle +|13\rangle-|15\rangle)+0(|2\rangle+|4\rangle +|6\rangle+|8\rangle+|10\rangle \nonumber \\
&+&|12\rangle+|14\rangle+|16\rangle)-0.2629(|5\rangle-|7\rangle+|9\rangle-|11\rangle)
\end{eqnarray}
\begin{eqnarray}
|k_x=0, n_y=4; \lambda = -0.382 \rangle &=& -0.1859(|1\rangle-|2\rangle +|3\rangle-|4\rangle+|13\rangle-|14\rangle+|15\rangle-|16\rangle) \nonumber \\ 
&-& 0.3008(|5\rangle-|6\rangle +|7\rangle-|8\rangle+|9\rangle-|10\rangle+|11\rangle-|12\rangle)
\end{eqnarray}
\begin{eqnarray}
|k_x=\pi, n_y=1; \lambda = 0.382 \rangle &=& 0.1859(|1\rangle+|2\rangle +|3\rangle+|4\rangle-|13\rangle-|14\rangle-|15\rangle-|16\rangle)-0.3008(|5\rangle+|6\rangle  \nonumber \\
&+&|7\rangle+|8\rangle-|9\rangle-|10\rangle-|11\rangle-|12\rangle)
\end{eqnarray}
\begin{eqnarray}
|k_x=\pi/2, n_y=3; \lambda = 0.618;1 \rangle &=& 0.4253(|1\rangle-|3\rangle -|13\rangle+|15\rangle)+0(|2\rangle+|4\rangle+|6\rangle+|8\rangle+|10\rangle+|12\rangle  \nonumber \\
&+& |14\rangle+|16\rangle)+0.2629(|5\rangle-|7\rangle-|9\rangle+|11\rangle)
\end{eqnarray}
\begin{eqnarray}
|k_x=3\pi/2, n_y=3; \lambda = 0.618;2 \rangle &=& 0(|1\rangle+|3\rangle +|5\rangle+|7\rangle+|9\rangle+|11\rangle+|13\rangle+|15\rangle)+0.4253(|2\rangle-|4\rangle \nonumber \\
&-&|14\rangle+|16\rangle)+0.2629(|6\rangle-|8\rangle-|10\rangle+|12\rangle)
\end{eqnarray}
\begin{eqnarray}
|k_x=\pi, n_y=2; \lambda = 1.382 \rangle &=& 0.3008(|1\rangle+|2\rangle +|3\rangle+|4\rangle+|13\rangle+|14\rangle+|15\rangle+|16\rangle)-0.1859(|5\rangle+|6\rangle \nonumber \\
&+&|7\rangle+|8\rangle+|9\rangle+|10\rangle+|11\rangle+|12\rangle)
\end{eqnarray}
\begin{eqnarray}
|k_x=\pi/2, n_y=4; \lambda = 1.618;1 \rangle &=& 0(|1\rangle+|3\rangle +|5\rangle+|7\rangle+|9\rangle+|11\rangle+|13\rangle+|15\rangle)+0.2629(|2\rangle-|4\rangle \nonumber \\
&+&|14\rangle-|16\rangle)+0.4253(|6\rangle-|8\rangle+|10\rangle-|12\rangle)
\end{eqnarray}
\begin{eqnarray}
|k_x=3\pi/2, n_y=4; \lambda = 1.618;2 \rangle &=& 0.2629(|1\rangle-|3\rangle +|13\rangle-|15\rangle)+0(|2\rangle+|4\rangle+|6\rangle+|8\rangle+|10\rangle \nonumber \\
&+&|12\rangle+|14\rangle+|16\rangle)+0.4253(|5\rangle-|7\rangle+|9\rangle-|11\rangle)
\end{eqnarray}
\begin{eqnarray}
|k_x=\pi, n_y=3; \lambda = 2.618 \rangle &=& 0.3008(|1\rangle+|2\rangle +|3\rangle+|4\rangle-|13\rangle-|14\rangle-|15\rangle-|16\rangle)+0.1859(|5\rangle+|6\rangle  \nonumber \\ 
&+&|7\rangle+|8\rangle-|9\rangle-|10\rangle-|11\rangle-|12\rangle)
\end{eqnarray}
\begin{eqnarray}
|k_x=\pi, n_y=4; \lambda = 3.618 \rangle &=& 0.1859(|1\rangle+|2\rangle +|3\rangle+|4\rangle+|13\rangle+|14\rangle+|15\rangle+|16\rangle)+0.3008(|5\rangle+|6\rangle \nonumber \\ 
&+&|7\rangle+|8\rangle+|9\rangle+|10\rangle+|11\rangle+|12\rangle)
\end{eqnarray}

\section{Finite-Size Scaling Analysis}
\label{FSS}
\subsection{Simple Hubbard Model: V=0}
\label{FSSV0}

\begin{figure*}[!htbp]
\centering
\includegraphics[width=0.98\textwidth]{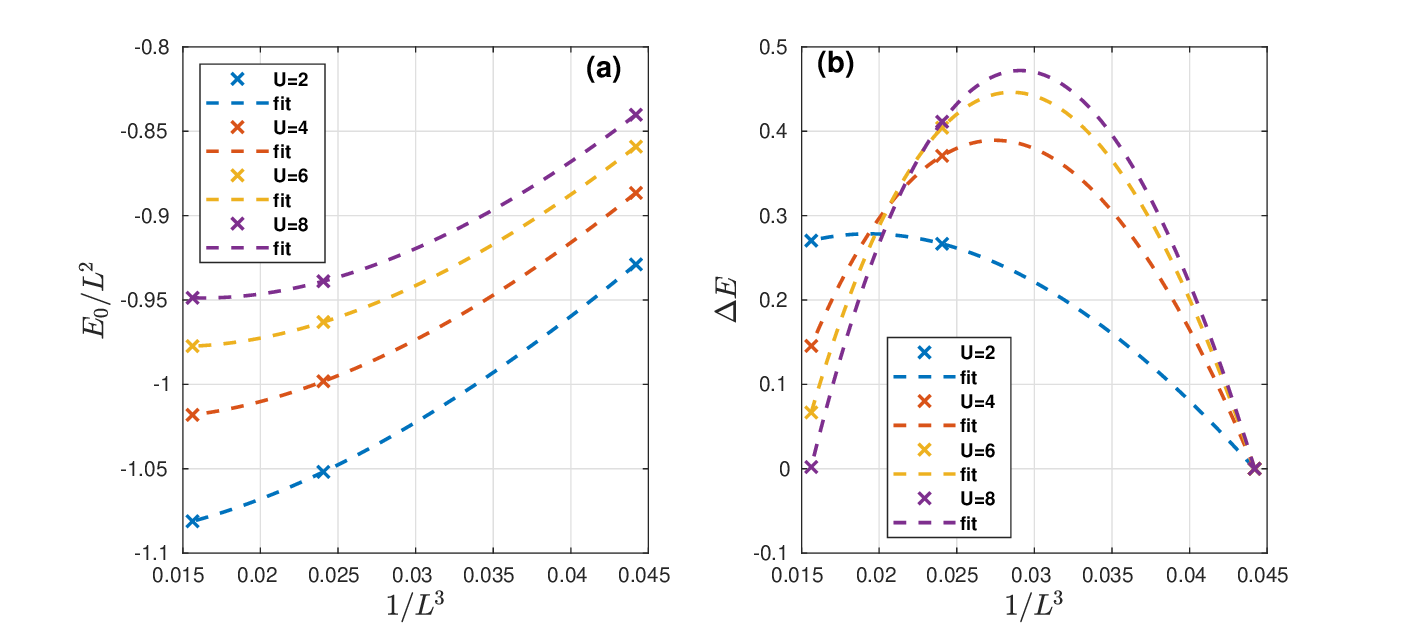}
\caption{Finite-size scaling analysis of (a) ground-state energy per site $E_0/L^2$ and (b) excitation gap $\Delta E = E_1 - E_0$ as functions of inverse system size $1/L^3$ for quarter-filled Hubbard clusters. Results are shown for $U = 2, 4, 6, 8$ using $2\times 4$, $3\times 4$ and $4\times 4$ cylindrical clusters. Dashed lines represent exact three-parameter fits to the scaling form $f(L) = f^\infty + x/L^3 + y/L^4$. The curvature in (b) underscores the significance of the $1/L^4$ subleading corrections in capturing non-monotonic finite-size effects as the system approaches the thermodynamic limit.}
\label{fig:E0gapfsstu}
\end{figure*}

To investigate the finite-size effects and the thermodynamic limit (TDL) behavior, we perform scaling analysis of the ground-state energy per site and the excitation gap for quarter-filled $2\times4$, $3\times4$ and $4\times4$ cylindrical Hubbard clusters. For these 2D systems, the effective linear size is defined as $L = \sqrt{M}$ \cite{Mingpu2016, Hao2024, Dian2024, Sandro2015, Tatche2010, Ahsan1994}, where $M$ is the number of sites. This corresponds to $1/L^3 \approx 0.0442$, $0.0241$ and $0.0156$ for $2\times4$, $3\times4$ and $4\times4$ clusters, respectively.

The ground-state energy per site ($E_0/L^2$) and the excitation gap ($\Delta E = E_1 - E_0$) are expected to follow the scaling forms:
\begin{equation}
\frac{E_0}{L^2} = e_0^\infty + \frac{x}{L^3} + \frac{y}{L^4} + \dots,
\label{scaling_E}
\end{equation}
\begin{equation}
\Delta E = \Delta E^\infty + \frac{x}{L^3} + \frac{y}{L^4} + \dots,
\label{scaling_Egap}
\end{equation}
where $e_0^\infty$ and $\Delta E^\infty$ are the values in the TDL ($1/L^3 \rightarrow 0$). The $1/L^3$ term represents the leading correction \cite{Mingpu2016, Hao2024, Herbert1989}, while the $1/L^4$ term accounts for higher-order subleading corrections \cite{Mingpu2016, Anders1997} in the corresponding quantities.

Figure \ref{fig:E0gapfsstu} presents the scaling results for $U=2, 4, 6,$ and $8$. We perform an exact fit to the three-parameter form ($e_0^\infty$ or $\Delta E^\infty$, $x$, $y$) \cite{Mingpu2016, Hao2024} as defined in Eqs. \ref{scaling_E} and \ref{scaling_Egap}, constrained by three available cluster sizes. The extracted parameters for the ground-state energy per site and the excitation gap are summarized in Tables \ref{tab:E0_scaling} and \ref{tab:Gap_scaling}, respectively. This procedure enables a more reliable extrapolation than a linear fit in $1/L^3$, which would implicitly neglect higher-order corrections (i.e., assume $y=0$).

In Fig. \ref{fig:E0gapfsstu}(a), $E_0/L^2$ exhibits a smooth, monotonic decrease as the system size increases ($1/L^3 \rightarrow 0$). The energy becomes more negative as we approach the TDL, with the fits providing a stable extrapolation for all $U$ values.

In Fig. \ref{fig:E0gapfsstu}(b), the excitation gap $\Delta E$ exhibits non-monotonic dependence on system size. For the smallest $2\times4$ cluster, $\Delta E$ remains nearly zero for all values of $U$. As the system size increases to $3\times4$ and $4\times4$ clusters (i.e., moving toward smaller $1/L^3$), the excitation gap first increases and then decreases as the TDL is approached. The inclusion of the $1/L^4$ term is essential to capture this curvature and provide a reliable extrapolation to the TDL.

\begin{table}[h]
\caption{Finite-size scaling parameters for the ground-state energy per site obtained from Eq.~(\ref{scaling_E}).}
\label{tab:E0_scaling}
\begin{ruledtabular}
\begin{tabular}{cccc}
$U$ & $e_0^\infty$ & $x$ & $y$ \\
\hline
2 & -1.070992 & -9.974031 & 37.304438 \\
\hline
4 & -0.977233 & -13.877778 & 45.057084 \\
\hline
6 & -0.919236 & -15.981920 & 49.050716 \\
\hline
8 & -0.875893 & -17.855587 & 52.785797 \\
\end{tabular}
\end{ruledtabular}
\end{table}

\begin{table}[h]
\caption{Finite-size scaling parameters for the excitation gap obtained from Eq.~(\ref{scaling_Egap}).}
\label{tab:Gap_scaling}
\begin{ruledtabular}
\begin{tabular}{cccc}
$U$ & $\Delta E^\infty$ & $x$ & $y$ \\
\hline
2 & -0.034383 & 64.712421 & -180.833840 \\
\hline
4 & -1.372678 & 256.755331 & -638.362337 \\
\hline
6 & -2.023927 & 346.258834 & -849.836537 \\
\hline
8 & -2.437914 & 399.978089 & -975.282360 \\
\end{tabular}
\end{ruledtabular}
\end{table}

\begin{figure}[t]
\centering
\includegraphics[width=0.50\textwidth]{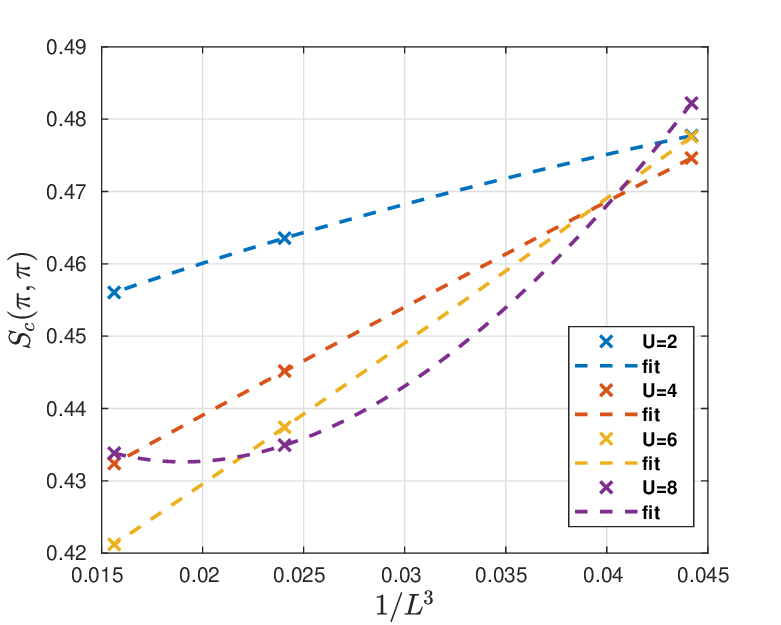}
\caption{Finite-size scaling of the charge structure factor $S_c(\pi,\pi)$ as a function of $1/L^3$ for $U = 2, 4, 6,$ and $8$.}
\label{fig:csffsstu}
\end{figure}

To characterize the charge degrees of freedom, we examine the finite-size scaling of the charge structure factor $S_c(\pi,\pi)$. Following the same approach as for the ground-state energy and excitation gap, we adopt the scaling form
\begin{equation}
S_c(\pi,\pi) = S_c^\infty + \frac{x}{L^3} + \frac{y}{L^4} + \dots,
\label{scaling_csf}
\end{equation}
where $S_c^\infty$ represents the value in the TDL. Figure \ref{fig:csffsstu} shows $S_c(\pi,\pi)$ as a function of $1/L^3$, together with the corresponding three-parameter fits. For $U=2, 4,$ and $6$, the data points are consistent with linear dependence on $1/L^3$, as is evident from small magnitude of the subleading coefficient $y$, shown in Table \ref{tab:csf_scaling}. At $U=8$, however, deviation from linear dependence emerges, signaling the increasing importance of subleading corrections. This change in scaling behavior indicates crossover from intermediate ($U=6$) to strong ($U=8$) coupling regime. The pronounced subleading corrections at $U=8$ indicates the suppression of charge fluctuations and emergence of spin ordering in the strong coupling regime, resulting in a more complex scaling behavior compared to low $U$ values, as the system approaches the TDL. 

\begin{table}[h]
\caption{Finite-size scaling parameters for the charge structure factor $S_c(\pi,\pi)$ obtained from the three-parameter fit described in Eq. (\ref{scaling_csf}).}
\label{tab:csf_scaling}
\begin{ruledtabular}
\begin{tabular}{cccc}
$U$ & $S_c^\infty$ & $x$ & $y$ \\
\hline
2 & 0.437534 & 1.848828 & -2.657823 \\
\hline
4 & 0.407315 & 1.800691 & -0.785417 \\
\hline
6 & 0.393059 & 1.534457 & 1.069737 \\
\hline
8 & 0.485824 & -11.165364 & 31.348997 \\
\end{tabular}
\end{ruledtabular}
\end{table}

\subsection{Extended Hubbard Model: $U=4$}
\label{FSSU4}
\begin{figure*}[!htbp]
\centering
\includegraphics[width=0.98\textwidth]{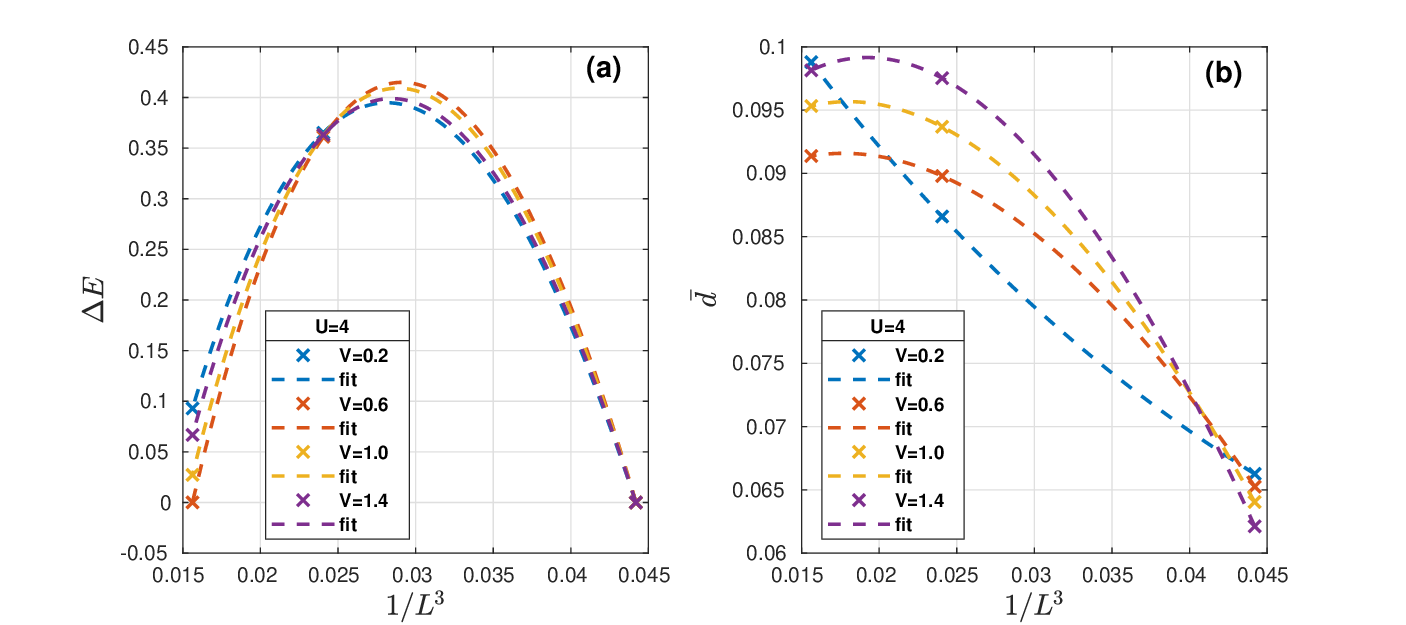}
\caption{Finite-size scaling analysis of (a) the excitation gap $\Delta E$ and (b) the average double occupancy $\bar{d}$ for the extended Hubbard model at fixed $U=4$, shown as function of $1/L^3$ for several values of $V$. Dashed lines represent exact fits to the scaling form $f(L) = f^\infty + x/L^3 + y/L^4$ using data from $2\times 4$, $3\times 4$ and $4\times 4$ clusters.}
\label{fig:Fig23}
\end{figure*}

To quantify the role of nearest-neighbor interaction $V$ on the excitation gap $\Delta E$ and the average double occupancy $\bar{d}$ at $U=4$, we carry out finite-size scaling analysis over $2\times4$, $3\times4$ and $4\times4$ clusters with cylindrical boundary conditions employing three-parameter fit that incorporates both leading $1/L^3$ and subleading $1/L^4$ corrections of the form
\begin{equation}
f(L) = f^\infty + \frac{x}{L^3} + \frac{y}{L^4},
\label{exact_fit_U4}
\end{equation}
where $f^\infty$ denotes the TDL value of the corresponding quantities. This form provides a controlled extrapolation by accounting for higher-order finite-size corrections, when only a limited number of system sizes are being investigated.

Figure \ref{fig:Fig23} summarizes the scaling behavior. The excitation gap $\Delta E$ exhibits a non-monotonic dependence on system size [Fig. \ref{fig:Fig23}(a)]. It is nearly zero for $2\times4$ cluster, increases to a maximum for $3\times4$ cluster, and then decreases for $4\times4$ cluster taking the TDL value as $1/L^3 \rightarrow 0$. This trend reflects the suppression of excitation gap with increasing system size, while the position of the maximum is governed primarily by finite-size effects.

The average double occupancy $\bar{d}$ [Fig. \ref{fig:Fig23}(b)] increases monotonically with system size (i.e., decreasing $1/L^3$) for all values of $V$. The curvature captured by the $1/L^4$ term, is most pronounced at small $V$ and becomes less prominent as $V$ increases. Concurrently, $\bar{d}$ decreases with increasing $V$ across all system sizes, and the fitted curves move closer together, indicating reduced finite-size dependence.

\subsection{Extended Hubbard Model: $U=10$}
\label{FSSU10}
\begin{figure*}[!htbp]
\centering
\includegraphics[width=0.98\textwidth]{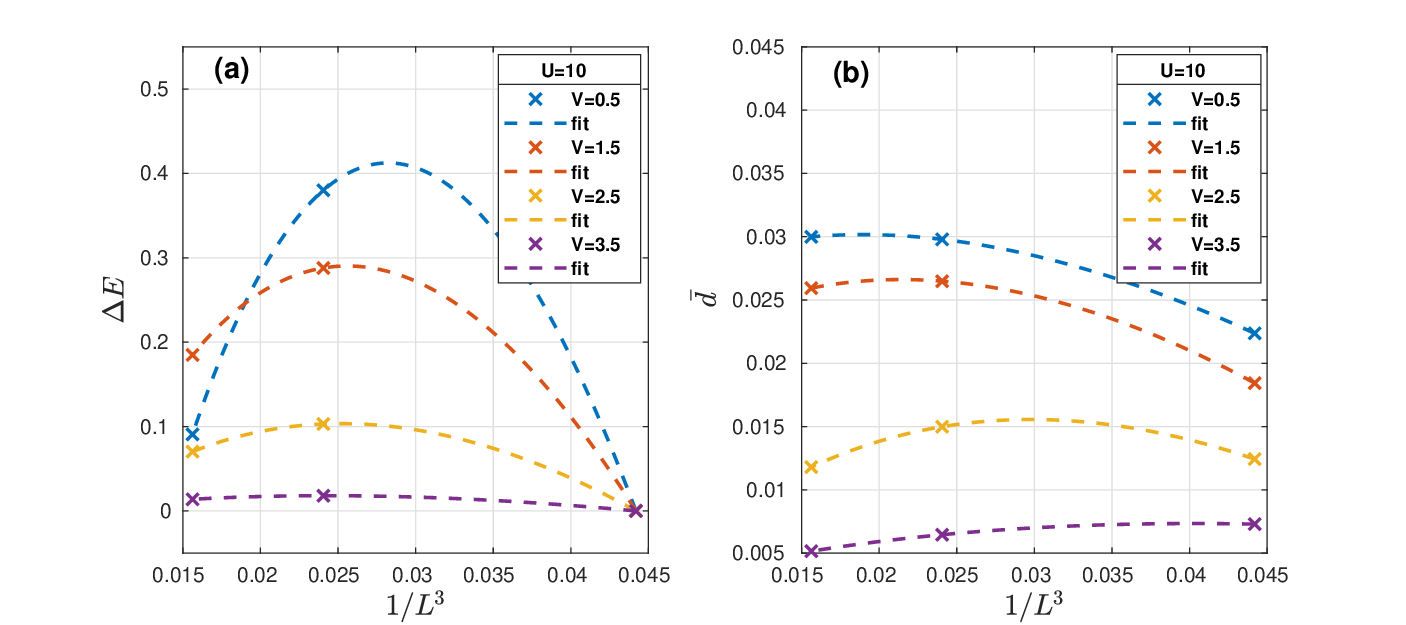}
\caption{Finite-size scaling analysis of (a) the excitation gap $\Delta E$ and (b) the average double occupancy $\bar{d}$ for the extended Hubbard model at fixed $U=10$, shown as function of $1/L^3$ for several values of $V$. Dashed lines represent exact fits to the scaling form $f(L) = f^\infty + x/L^3 + y/L^4$ using data from $2\times 4$, $3\times 4$ and $4\times 4$ clusters.}
\label{fig:Fig24}
\end{figure*}

Finally, we examine the finite-size scaling behavior of the excitation gap $\Delta E$ and the average double occupancy $\bar{d}$ at $U=10$ for $V=0.5, 1.5, 2.5$ and $3.5$, using the same three-parameter scaling form [Eq. \ref{exact_fit_U4}] as in $U=4$ case. As shown in Fig. \ref{fig:Fig24}(a), $\Delta E$ exhibits non-monotonic size dependence, reaching a maximum at $3\times4$ cluster and subsequently decreasing as $1/L^3 \rightarrow 0$. The overall magnitude of the gap is progressively suppressed with increasing $V$, nearly vanishing at $V=3.5$.

In Fig. \ref{fig:Fig24}(b), the scaling of $\bar{d}$ shows weak curvature, particularly at small $V$, which is captured by the inclusion of subleading corrections. With increasing $V$, $\bar{d}$ decreases substantially and the size dependence becomes increasingly flat, approaching a small, nearly size-independent value ($\bar{d}\sim 0.005$) at $V=3.5$. 

Thus, compared to the $U=4$ case, the strong-coupling case with $U=10$ is characterized by rapid suppression of the average double occupancy $\bar{d}$, reflecting effects of enhanced correlations and faster convergence towards a regime with strongly suppressed charge fluctuations. 
\section{Explained variance ratios for charge correlations on the \(3\times4\) cluster}
\label{pcdata}
Table \ref{tab:appD_pca_charge} lists the explained variance ratios \(\tilde{\lambda}_k\) for the charge correlation matrix for the first six principal components, together with the sum \(\tilde{\lambda}_1+\tilde{\lambda}_2\), for the quarter‑filled simple Hubbard model (\(V=0\)) on \(3\times4\) cylindrical cluster. The data correspond to PCA analysis discussed in Sec. \ref{sec:pcashm} and in Fig. \ref{fig:3x4pcav0}.

\begin{table}[htbp]
\centering
\caption{Explained variance ratios \(\tilde{\lambda}_k\) (\(k=1,\cdots ,6\)) and \(\tilde{\lambda}_1+\tilde{\lambda}_2\) for charge correlations at different \(U\) values.}
\label{tab:appD_pca_charge}
\begin{ruledtabular}
\begin{tabular}{cccccccc}
\(U\) & \(\tilde{\lambda}_1\) & \(\tilde{\lambda}_2\) & \(\tilde{\lambda}_3\) & \(\tilde{\lambda}_4\) & \(\tilde{\lambda}_5\) & \(\tilde{\lambda}_6\) & \(\tilde{\lambda}_1\)+\(\tilde{\lambda}_2\) \\
\hline
0  & 0.2091 & 0.1221 & 0.1192 & 0.1156 & 0.1148 & 0.0791 & 0.3312 \\
\hline
2  & 0.1543 & 0.1414 & 0.1412 & 0.1385 & 0.0868 & 0.0748 & 0.2957 \\
\hline
4  & 0.1650 & 0.1421 & 0.1411 & 0.1398 & 0.0881 & 0.0745 & 0.3071 \\
\hline
6  & 0.1742 & 0.1424 & 0.1407 & 0.1406 & 0.0880 & 0.0737 & 0.3166 \\
\hline
8  & 0.1820 & 0.1424 & 0.1412 & 0.1401 & 0.0873 & 0.0729 & 0.3244 \\
\hline
10 & 0.1886 & 0.1420 & 0.1415 & 0.1397 & 0.0864 & 0.0722 & 0.3305 \\
\end{tabular}
\end{ruledtabular}
\end{table}

\section{Spatial Distribution of Subleading Principal Component Eigenvectors}
\label{subpcaeigvec}
To illustrate the spatial distribution of subleading principal components, we show representative higher-order eigenvectors ($w_2$, $w_3$) of the charge-correlation matrix for $4\times4$ cluster at different interaction strengths ($U,V$) in Fig.~\ref{fig:Fig25}. These subleading components exhibit short-range or irregular modulations and are small compared to the leading principal component eigenvector $w_1$, confirming that the subleading components do not alter the dominant charge-ordering tendency discussed in the main text.

\begin{figure*}[!htbp]
\centering
\includegraphics[scale=0.35]{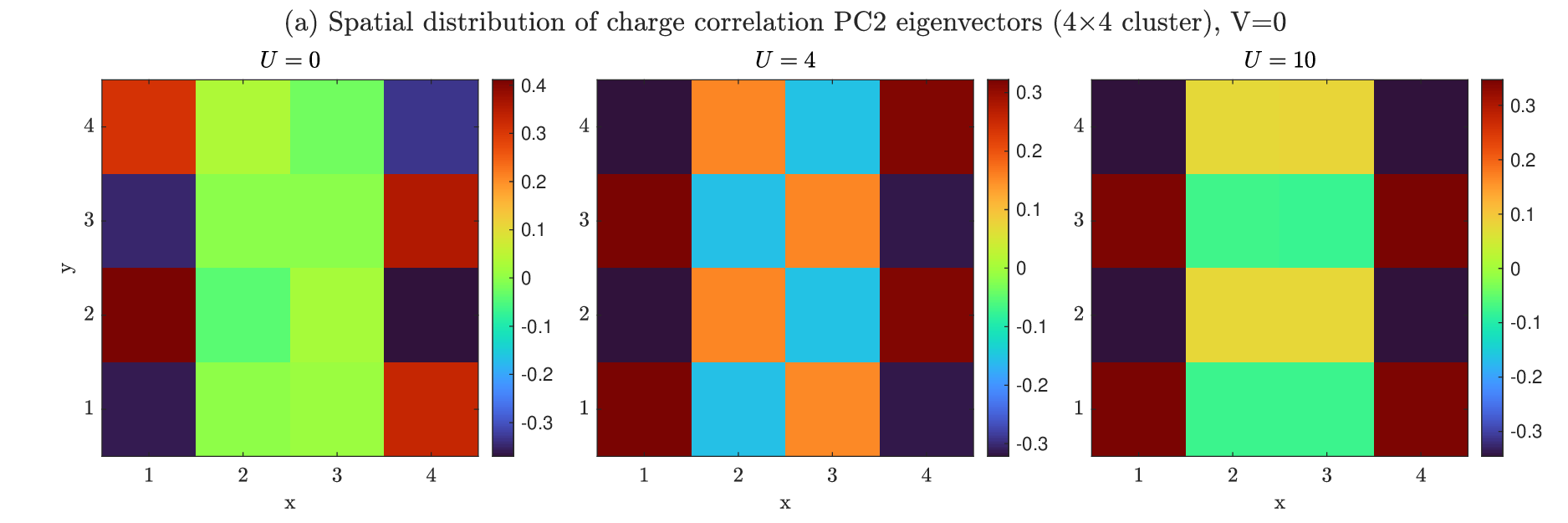}
\quad
\includegraphics[scale=0.35]{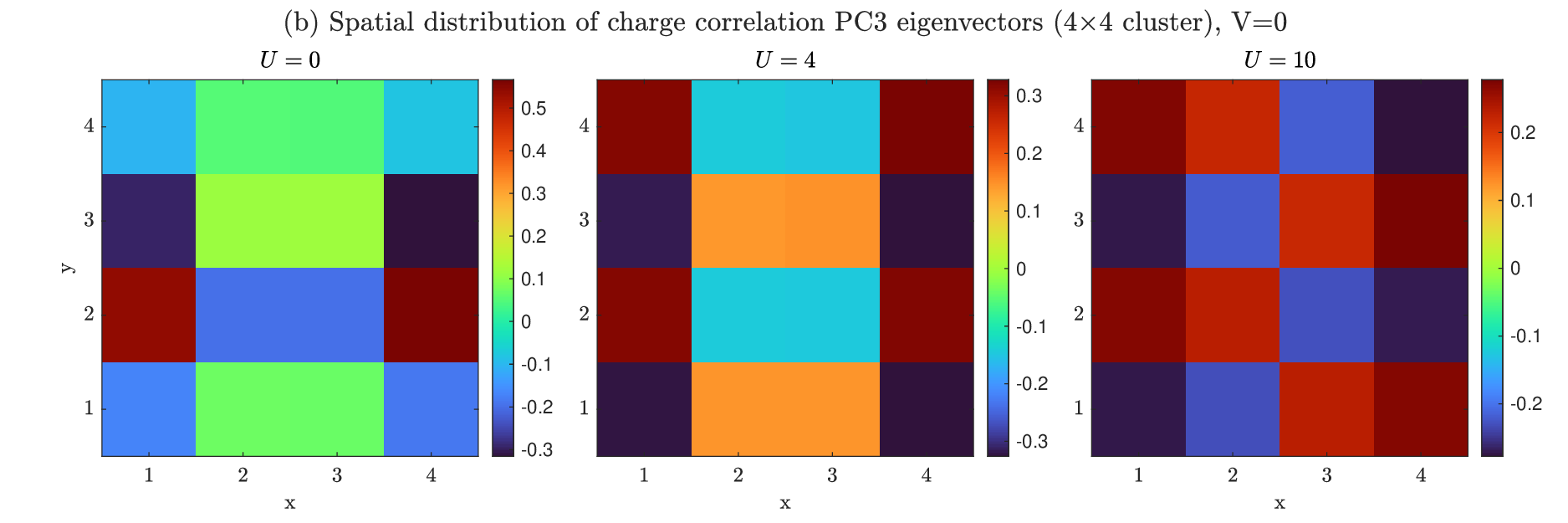}
\quad
\includegraphics[scale=0.35]{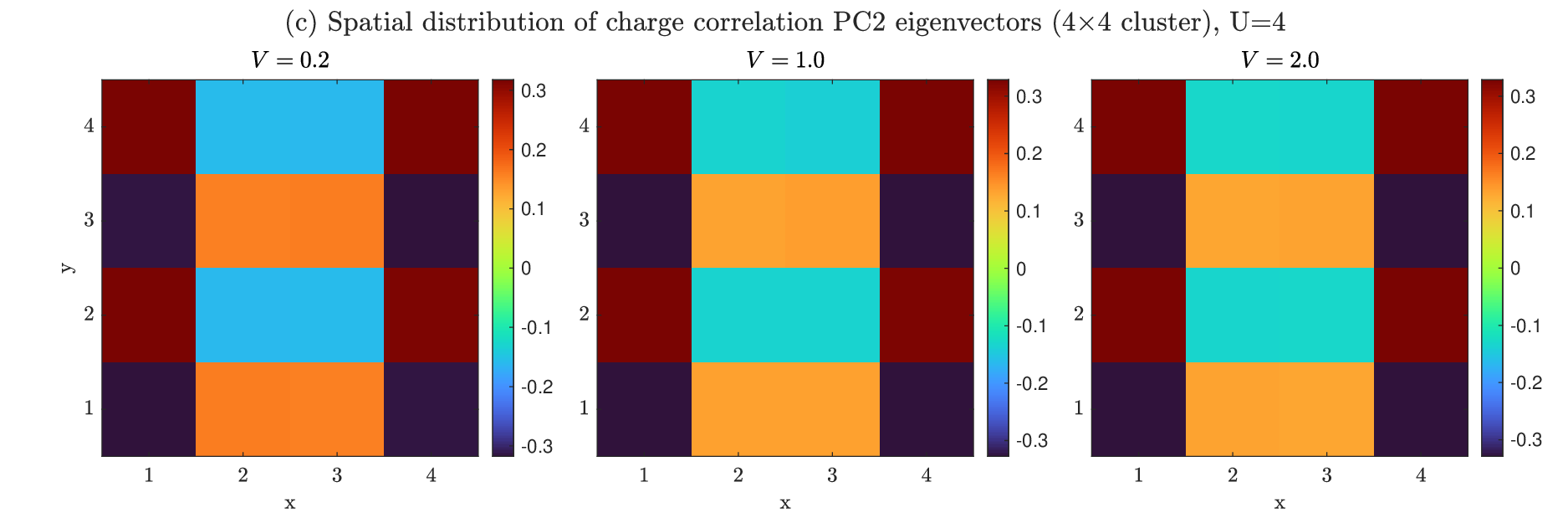}
\quad
\includegraphics[scale=0.35]{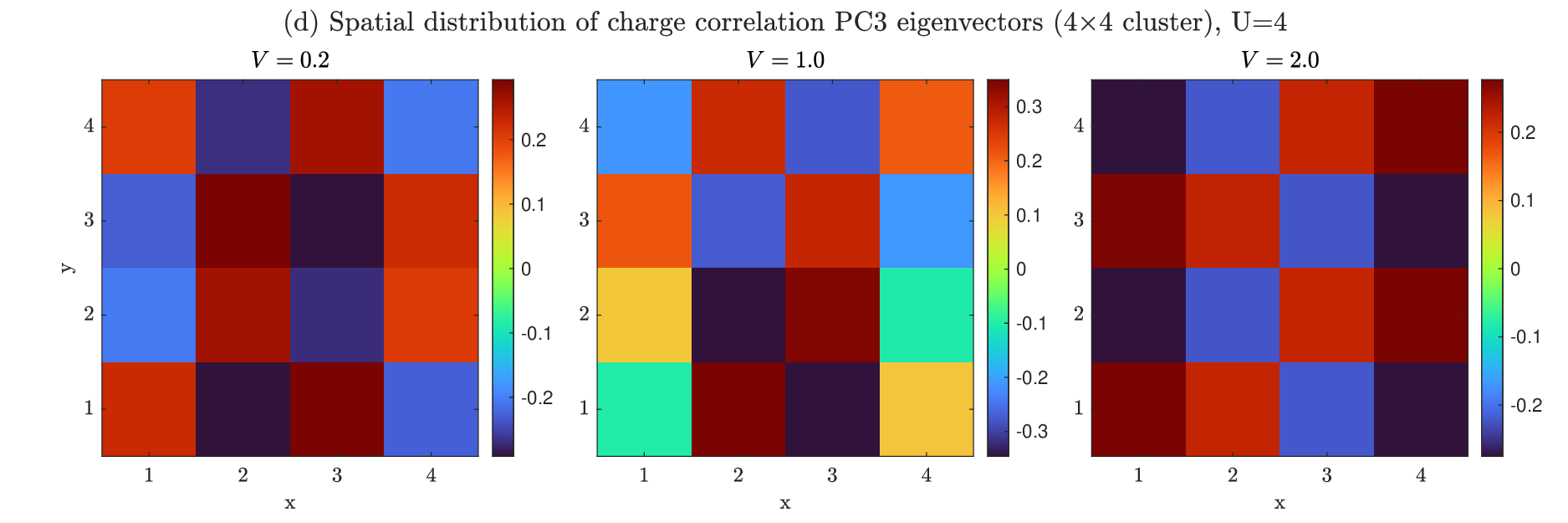}
\caption{Representative subleading eigenvectors ($w_2$, $w_3$) of the charge-correlation matrix for $4\times4$ cluster at different interaction strengths. Panels (a,b) correspond to $V=0$ for varying $U$, while (c,d) show evolution with $V$ at fixed $U=4$.}
\label{fig:Fig25}
\end{figure*}

\end{widetext}
\clearpage

\bibliography{main}

\end{document}